\documentclass[]{pasj01}
\begin{document} 
\Received{}%{yyyy/mm/dd}
\Accepted{}%{yyyy/mm/dd}
%\Published{yyyy/mm/dd}

\title{Subaru High-$z$ Exploration of Low-Luminosity Quasars (SHELLQs). II. Discovery of 32 Quasars and Luminous Galaxies at $5.7 < z \le 6.8$}

%%% begin:list of authors
% Do NOT capitalize all letters in "textsc".
\author{Yoshiki Matsuoka\altaffilmark{1,2,3}
%\thanks{Example: Present Address is xxxxxxxxxx}
}
\email{yk.matsuoka@cosmos.ehime-u.ac.jp}

\author{Masafusa Onoue\altaffilmark{1, 2}}
\author{Nobunari Kashikawa\altaffilmark{1,2}}
\author{Kazushi Iwasawa\altaffilmark{4}}
\author{Michael A. Strauss\altaffilmark{5}}
\author{Tohru Nagao\altaffilmark{3}}
\author{Masatoshi Imanishi\altaffilmark{1, 2, 6}}
\author{Chien-Hsiu Lee\altaffilmark{6}}
\author{Masayuki Akiyama\altaffilmark{7}}
\author{Naoko Asami\altaffilmark{8}}
\author{James Bosch\altaffilmark{5}}
\author{S\'{e}bastien Foucaud}
\author{Hisanori Furusawa\altaffilmark{1}}
\author{Tomotsugu Goto\altaffilmark{10}}
\author{James E. Gunn\altaffilmark{5}}
\author{Yuichi Harikane\altaffilmark{11, 12}}
\author{Hiroyuki Ikeda\altaffilmark{1}}
\author{Takuma Izumi\altaffilmark{1}}
\author{Toshihiro Kawaguchi\altaffilmark{13}}
\author{Satoshi Kikuta\altaffilmark{2}}
\author{Kotaro Kohno\altaffilmark{14, 15}}

\author{Yutaka Komiyama\altaffilmark{1, 2}}
\author{Robert H. Lupton\altaffilmark{5}}
\author{Takeo Minezaki\altaffilmark{14}}
\author{Satoshi Miyazaki\altaffilmark{1, 2}}
\author{Tomoki Morokuma\altaffilmark{14}}
\author{Hitoshi Murayama\altaffilmark{16}}
\author{Mana Niida\altaffilmark{17}}
\author{Atsushi J. Nishizawa\altaffilmark{18}}
\author{Masamune Oguri\altaffilmark{12,15,16}}
\author{Yoshiaki Ono\altaffilmark{11}}
\author{Masami Ouchi\altaffilmark{11,16}}
\author{Paul A. Price\altaffilmark{5}}
\author{Hiroaki Sameshima\altaffilmark{19}}
\author{Andreas Schulze\altaffilmark{1}}
\author{Hikari Shirakata\altaffilmark{20}}
\author{John D. Silverman\altaffilmark{16}}
\author{Naoshi Sugiyama\altaffilmark{16, 21}}
\author{Philip J. Tait\altaffilmark{6}}
\author{Masahiro Takada\altaffilmark{16}}
\author{Tadafumi Takata\altaffilmark{1, 2}}
\author{Masayuki Tanaka\altaffilmark{1, 2}}
\author{Ji-Jia Tang\altaffilmark{22}}
\author{Yoshiki Toba\altaffilmark{22}}
\author{Yousuke Utsumi\altaffilmark{23}}
\author{Shiang-Yu Wang\altaffilmark{22}}

\altaffiltext{1}{National Astronomical Observatory of Japan, Mitaka, Tokyo 181-8588, Japan.}
\altaffiltext{2}{Department of Astronomical Science, Graduate University for Advanced Studies (SOKENDAI), Mitaka, Tokyo 181-8588, Japan.}
\altaffiltext{3}{Research Center for Space and Cosmic Evolution, Ehime University, Matsuyama, Ehime 790-8577, Japan.}
\altaffiltext{4}{ICREA and Institut de Ci{\`e}ncies del Cosmos, Universitat de Barcelona, IEEC-UB, Mart{\'i} i Franqu{\`e}s, 1, 08028 Barcelona, Spain.}
\altaffiltext{5}{Princeton University Observatory, Peyton Hall, Princeton, NJ 08544, USA.}
\altaffiltext{6}{Subaru Telescope, Hilo, HI 96720, USA.}
\altaffiltext{7}{Astronomical Institute, Tohoku University, Aoba, Sendai, 980-8578, Japan.}
\altaffiltext{8}{Japan Professional School of Education, Chiyoda, Tokyo 101-0041, Japan.}
\altaffiltext{9}{Department of Physics and Astronomy, Shanghai JiaoTong University, Shanghai 200240, China.}
\altaffiltext{10}{Institute of Astronomy and Department of Physics, National Tsing Hua University, Hsinchu 30013, Taiwan.}
\altaffiltext{11}{Institute for Cosmic Ray Research, The University of Tokyo, Kashiwa, Chiba 277-8582, Japan}
\altaffiltext{12}{Department of Physics, Graduate School of Science, The University of Tokyo, Bunkyo, Tokyo 113-0033, Japan}
%\altaffiltext{13}{Department of Liberal Arts and Sciences, Sapporo Medical University, Chuo, Sapporo 060-8556, Japan.}
\altaffiltext{13}{Department of Economics, Management and Information Science, Onomichi City University, Onomichi, Hiroshima 722-8506, Japan.}
\altaffiltext{14}{Institute of Astronomy, The University of Tokyo, Mitaka, Tokyo 181-0015, Japan.}
\altaffiltext{15}{Research Center for the Early Universe, University of Tokyo, Tokyo 113-0033, Japan.}
\altaffiltext{16}{Kavli Institute for the Physics and Mathematics of the Universe, WPI, The University of Tokyo,Kashiwa, Chiba 277-8583, Japan.}
\altaffiltext{17}{Graduate School of Science and Engineering, Ehime University, Matsuyama, Ehime 790-8577, Japan.}
\altaffiltext{18}{Institute for Advanced Research, Nagoya University, Furo-cho, Chikusa-ku, Nagoya 464-8602, Japan.}
\altaffiltext{19}{Koyama Astronomical Observatory, Kyoto-Sangyo University, Kita, Kyoto, 603-8555, Japan.}
\altaffiltext{20}{Department of Cosmosciences, Graduates School of Science, Hokkaido University, N10 W8, Kitaku, Sapporo 060-0810, Japan.}
\altaffiltext{21}{Graduate School of Science, Nagoya University, Furo-cho, Chikusa-ku, Nagoya 464-8602, Japan.}
\altaffiltext{22}{Institute of Astronomy and Astrophysics, Academia Sinica, Taipei, 10617, Taiwan.}
\altaffiltext{23}{Hiroshima Astrophysical Science Center, Hiroshima University, Higashi-Hiroshima, Hiroshima 739-8526, Japan.}

%%% end:list of authors

%% `\KeyWords{}' always has to be placed before `\maketitle'.
 %Do NOT move this preamble from here!

\maketitle

\begin{abstract}
We present spectroscopic identification of 32 new quasars and luminous galaxies discovered at $5.7 < z \le 6.8$.
This is the second in a series of papers presenting the results of the Subaru High-$z$ Exploration of Low-Luminosity Quasars (SHELLQs) project, which 
exploits the deep multi-band imaging data produced by the Hyper Suprime-Cam (HSC) Subaru Strategic Program survey.
The photometric candidates were selected by a Bayesian probabilistic algorithm, and then observed with spectrographs on the Gran Telescopio Canarias and the Subaru Telescope.
Combined with the sample presented in the previous paper, we have now identified 64 HSC sources over about 430 deg$^2$, which include
33 high-$z$ quasars, 14 high-$z$ luminous galaxies, 2 [O\emissiontype{III}] emitters at $z \sim 0.8$, and 15 Galactic brown dwarfs.
The new quasars have considerably lower luminosity ($M_{1450} \sim -25$ to $-22$ mag) than most of the previously known high-$z$ quasars.
Several of these quasars have luminous ($> 10^{43}$ erg s$^{-1}$) and narrow ($< 500$ km s$^{-1}$) Ly$\alpha$ lines, and also a possible mini broad absorption line system of N\emissiontype{V} $\lambda$1240 
in the composite spectrum, which clearly separate them from typical quasars.
On the other hand, the high-$z$ galaxies have extremely high luminosity ($M_{1450} \sim -24$ to $-22$ mag) compared to other galaxies found at similar redshift.
With the discovery of these new classes of objects, we are opening up new parameter spaces in the high-$z$ Universe.
Further survey observations and follow-up studies of the identified objects, including the construction of the quasar luminosity function at $z \sim 6$, are ongoing.
\end{abstract}

\section{Introduction}

High-$z$ quasars are a unique and useful probe of the early Universe in many aspects. 
The progress of cosmic reionization has been measured by the strength of H \emissiontype{I} absorption in background quasar spectra, which is very sensitive to 
the neutral fraction of the intergalactic medium (IGM; \cite{gunn65}; \cite{fan06araa}).
Stringent constraints on the seed population and initial growth of supermassive black holes (SMBHs) can be obtained from their mass function, in particular
their maximum mass, as a function of redshift (e.g., \cite{volonteri12}; \cite{ferrara14}; \cite{madau14}).
We can also learn about the formation of their host galaxies, which is thought to have happened in the highest density peaks of the underlying dark matter distribution
in the early phase of cosmic history.

There has been great progress in the quest for high-$z$ quasars\footnote{Hereafter, ``high-$z$" denotes $z > 5.7$, where quasars are observed as $i$-band 
dropouts in the Sloan Digital Sky Survey (SDSS) filter system \citep{fukugita96}.} in the last two decades.
This progress was made possible by the advent of wide-field ($\ge$ 1000 deg$^2$) surveys in the optical or near-infrared (IR) bands, such as
SDSS \citep{york00}, the Canada-France-Hawaii Telescope Legacy Survey, the United Kingdom Infrared Telescope (UKIRT) Infrared Deep Sky Survey 
(UKIDSS; \cite{lawrence07}), the Panoramic Survey Telescope \& Rapid Response System 1 (Pan-STARRS1; \cite{kaiser10}) 3$\pi$ survey,
the Dark Energy Survey \citep{des16}, and the Visible and Infrared Survey Telescope for Astronomy (VISTA) Kilo-degree Infrared Galaxy (VIKING).
High-$z$ quasar discoveries from the above and other projects are reported in 
Fan et al. (2000, 2001a, 2003, 2004, 2006b), Jiang et al. (2008, 2009, 2015, 2016), %\citet{fan00, fan01, fan03, fan04, fan06, jiang08, jiang09, jiang15}.
Willott et al. (2005, 2007, 2009, 2010ab), %\citet{willott05, willott07, willott09, willott10a,willott10} 
\citet{mortlock11}, 
Ba{\~n}ados et al. (2014, 2016), %\citet{banados14, banados16}
Reed et al. (2015, 2017), %\citet{reed15, reed17}
Venemans et al. (2013, 2015ab), %\citet{venemans13, venemans15a, venemans15b}
\citet{goto06}, \citet{carnall15}, \citet{kashikawa15}, \citet{kim15}, \citet{wu15}, and \citet{wang17}.

More than 100 high-$z$ quasars are known today \citep{banados16}, with the most distant object found at $z = 7.085$ \citep{mortlock11}.
However, most of these quasars have redshifts $z < 6.5$ and absolute magnitudes $M_{1450} < -24$ mag, while higher redshifts and fainter magnitudes are still poorly explored.
%The known bright high-$z$ quasars 
The known quasars must be just the tip of an iceberg predominantly comprised of faint quasars and AGNs, which may be a significant
contributor to reionization, and may represent the more typical mode of SMBH growth in the early Universe.

This paper is the second in a series presenting the results of the Subaru High-$z$ Exploration of Low-Luminosity Quasars (SHELLQs) project, which 
%is the first 1,000-deg$^2$ class survey for high-$z$ quasars with a 8-m class telescope.
exploits multi-band photometry data produced by the Hyper Suprime-Cam (HSC) Subaru Strategic Program (SSP) survey.
The first results were presented in Matsuoka et al. (2016; hereafter Paper I), where we described the motivation and strategy of the project in detail, 
as well as the discovery of 15 quasars and luminous galaxies at $5.7 < z < 6.9$ from the initial 80 deg$^2$ of the survey.
In the present paper, we report the discovery of an additional 24 quasars and 8 luminous galaxies at $z > 5.7$, from about 430 deg$^2$ (including the previous 80 deg$^2$) of the survey.
The spectroscopic follow-up campaign on the present survey area is still ongoing, whose results will be presented in forthcoming papers.
We are also working to construct quasar luminosity function at $z \sim 6$, which will be presented in a separate paper.

This paper is organized as follows.
We briefly describe the photometric candidate selection in \S \ref{sec:selection}, while a more complete description is found in Paper I.
The spectroscopic follow-up observations are described in \S \ref{sec:spectroscopy}.
The quasars and galaxies we have discovered are presented and discussed in \S \ref{sec:results}.
The summary appears in \S \ref{sec:summary}.
We adopt the cosmological parameters $H_0$ = 70 km s$^{-1}$ Mpc$^{-1}$, $\Omega_{\rm M}$ = 0.3, and $\Omega_{\rm \Lambda}$ = 0.7.
All magnitudes in the optical and NIR bands are presented in the AB system \citep{oke83}.
Magnitudes refer to point spread function (PSF) magnitudes (see \S \ref{subsec:hscsurvey}) 
corrected for Galactic extinction \citep{schlegel98},
unless otherwise noted.
In what follows, we refer to $z$-band magnitudes with the AB subscript (``$z_{\rm AB}$"), while redshift $z$ appears without a subscript.

\section{Photometric Candidate Selection \label{sec:selection}}

Our quasar candidates were selected in essentially the same way as in Paper I, so we only briefly describe the relevant procedure here, highlighting 
a few minor changes we made.
The reader is referred to Paper I for a more complete description of our selection.

\subsection{The Subaru HSC-SSP Survey \label{subsec:hscsurvey}}

The SHELLQs project is based on the imaging data collected by the Subaru SSP survey with the HSC  (\cite{miyazaki12}, Miyazaki et al., in prep.), 
a wide-field camera installed on the Subaru 8.2 m telescope. % on the summit of Maunakea.
HSC has a nearly circular field of view of 1$^\circ$.5 diameter, covered by 116 2K $\times$ 4K Hamamatsu fully depleted CCDs, with the pixel scale of 0\arcsec.17.
The survey has three layers:
the Wide layer is observing 1400 deg$^2$ mostly along the celestial equator, with the 5$\sigma$ target depths of 
($g_{\rm AB}$, $r_{\rm AB}$, $i_{\rm AB}$, $z_{\rm AB}$, $y_{\rm AB}$) = (26.5, 26.1, 25.9, 25.1, 24.4) mag measured in 2\arcsec.0 apertures,
while the Deep and the UltraDeep layers are observing smaller areas (27 and 3.5 deg$^2$) down to deeper limiting magnitudes ($r_{\rm AB}$ = 27.1 and 27.7 mag, respectively).
A full description of the survey may be found in Aihara et al. (in prep.). %, while the relevant information to the SHELLQs project is summarized in Paper I.
The first public data release (DR1) took place in 2017 February, which includes the data taken in the first 1.7 years (2014 March to 2015 November) of the survey,
covering 108 deg$^2$ of the Wide layer and the Deep and UltraDeep layers to intermediate depths \citep{aihara17}.
The median seeing during the above observing period was 0\arcsec.5 -- 0\arcsec.8, depending on filter. % and observed field.
The DR1 Wide layer reaches limiting magnitudes consistent with the above target values, while we are still accumulating exposures in the Deep and UltraDeep layers
to reach the final target depths.

The results presented in this paper were drawn from imaging data observed before 2016 April, i.e., a newer dataset than contained in DR1.
We used about 430 deg$^2$ of the Wide layer, in which we have more than one exposure in the $i$, $z$, and $y$-bands.\footnote{
The quasar selection presented in this work was not restricted to the areas observed to the planned full depth.
We didn't use the higher-quality data from the Deep and UltraDeep layers, which are available over a small fraction of the Wide field.}
%, we used only the Wide-layer data for the quasar selection presented in this work.}
%; we describe later how this effective observed area was estimated.
%The observed fields are mostly along the celestial equator.
The total integration time per pointing in each of the ($i$, $z$, $y$) bands in the Wide layer is 20 minutes, divided into six individual exposures with different dither positions.
Data reduction was performed with the dedicated pipeline {\tt hscPipe} (version 4.0.1 and 4.0.2; Bosch et al., in prep.) derived from
the Large Synoptic Survey Telescope software pipeline \citep{juric15}, for all the standard procedures
including bias subtraction, flat fielding with dome flats, stacking, astrometric and photometric calibrations, as well as source detection and measurements.
The astrometric and photometric calibrations are tied to the Pan-STARRS1 system \citep{schlafly12,tonry12,magnier13, chambers16, flewelling16}.
We use the PSF magnitude ($m_{\rm AB}$) and the CModel magnitude ($m_{\rm CModel, AB}$),
which are measured by fitting the PSF models and two-component, PSF-convolved galaxy models to the source profile, respectively \citep{abazajian04}.
These magnitudes have been corrected for Galactic extinction \citep{schlegel98}.

\subsection{Candidate Selection \label{subsec:selection}}

\begin{figure}
 \begin{center}
  \includegraphics[width=8cm]{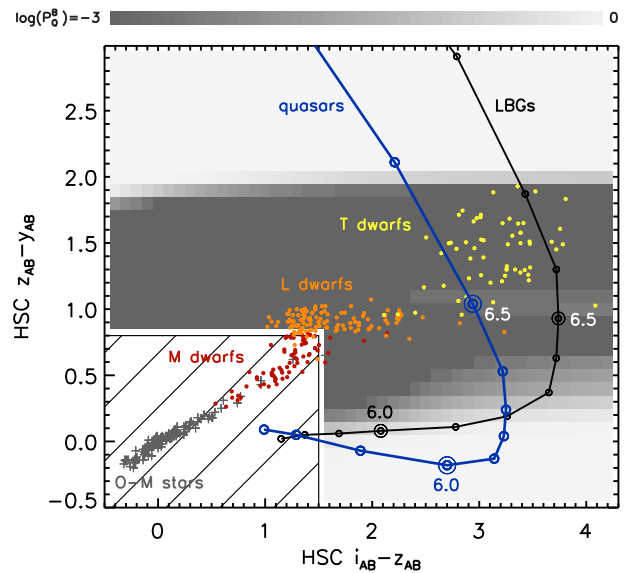} 
 \end{center}
\caption{HSC $i_{\rm AB} - z_{\rm AB}$ and $z_{\rm AB} - y_{\rm AB}$ colors of high-$z$ quasars (blue line) and galaxies (black line), as well as Galactic stars and brown dwarfs (crosses and dots).
The models and references used to compute these colors are described in Paper I.
%The SED models for quasars and brown dwarfs are described in \S \ref{subsec:mainsample}.
%The galaxy colors are calculated with the spectral templates taken from \citet{gonzalez12}, while the colors of O -- M stars are computed with the \citet{pickles98} library.
The open circles along the blue and black lines represent redshifts in steps of 0.1, with $z = 6.0$ and $6.5$ marked by the larger circles.
The hatched area in lower left indicates the color space excluded from the SSP database query in the first step of the quasar selection (\S \ref{subsec:selection}).
The grey scale represents the Bayesian quasar probability $P_{\rm Q}^{\rm B}$ (the color bar is found at the top) over this plane,
%Note that the $P_{\rm Q}^{\rm B}$ distribution changes in response to source and limiting magnitudes;
%here we assume 
for a source magnitude of $z_{\rm AB} = 24.0$ mag and 5$\sigma$ limiting magnitudes of ($i_{\rm AB}$, $z_{\rm AB}$, $y_{\rm AB}$) = (26.5, 25.5, 25.0) mag.
%The grey scale represents the distribution of $P_{\rm Q}^{\rm B}$ from 0 (black) to 1 (white).
Galaxy models are not included in our algorithm at present.
}\label{fig:colordiagram}
\end{figure}

We start our candidate selection from the HSC-SSP source catalog over the Wide layer. %, included in the S16A internal data release.
The initial query from the database is as follows:
\begin{eqnarray}
(z_{\rm AB} < 24.5\ {\rm and}\ \sigma_z < 0.155\ {\rm and}\ i_{\rm AB} - z_{\rm AB} > 1.5 \nonumber \\
 {\rm and}\ z_{\rm AB} - z_{\rm CModel, AB} < 0.15 )
\label{eq:query1}
 \end{eqnarray}
 or
 \begin{eqnarray}
(y_{\rm AB} < 24.0\ {\rm and}\ \sigma_y < 0.155\ {\rm and}\ z_{\rm AB} - y_{\rm AB} > 0.8 \nonumber \\
 {\rm and}\ y_{\rm AB} - y_{\rm CModel, AB} < 0.15 ) .
\label{eq:query2}
\end{eqnarray}
We reject objects with the critical quality flags detailed in Paper I.
Throughout this paper, ($i_{\rm AB}$, $z_{\rm AB}$, $y_{\rm AB}$) refer to PSF magnitudes, and ($\sigma_i$, $\sigma_z$, $\sigma_y$) refer to their errors.
The conditions of Equation \ref{eq:query1} select $i$-band dropouts at $z \sim 6$, while those of Equation \ref{eq:query2} select $z$-band dropouts at $z \sim 7$.
The color cuts are used to remove relatively blue stars with O to early-M spectral types (see Figure \ref{fig:colordiagram}),
while the difference between the PSF and CModel magnitudes is used to exclude extended sources.
We changed the extendedness cut from $m_{\rm AB} - m_{\rm CModel, AB} < 0.30$ in Paper I to $< 0.15$ in the present work, so that galaxies are removed more efficiently;
%in order to be more efficient in removing extended sources;
we discuss this issue later in this paper.
After the database query, we further remove low-$z$ interlopers with more than $3\sigma$ detection in the $g$ or $r$ band.

Next, an automatic image checking procedure %with Source Extractor (\cite{bertin96}; see Paper I) 
is run on stacked images and per-visit images (i.e., images of individual exposures before stacking)
of the above sources. %, which eliminates cosmic rays, moving or transient sources, image artifacts, and other false candidates.
This procedure uses Source Extractor \citep{bertin96} in the double-image mode, with the stacked image as the detection reference.
If any of the per-visit photometric measurements deviate by more than three times the measurement error from the stacked photometry, 
the source is eliminated.
We also reject sources with too compact, diffuse, or elliptical profiles to be celestial point sources %, with the Source Extractor measurements
on the stacked images.
%We defined conservative rejection criteria, so that we would not throw away any possible quasars at this stage.
The eliminated sources are mostly cosmic rays, moving or transient sources, and image artifacts.

The red HSC sources selected above are matched to the UKIDSS and VIKING catalogs within 1\arcsec.0, %in the overlapping survey area, 
which provides $J$, $H$, and $K$-band photometry.
The present work utilizes the UKIDSS data release 10 and the VIKING data release 4.
The UKIDSS data cover almost the entire HSC survey fields, while the available VIKING data cover 30--40 \% of the HSC survey footprint used in this work.

We use a template-based Bayesian probabilistic method, which selects quasar candidates from the above sample of red HSC sources.
Similar methods of spectral energy distribution (SED) fitting have been used in other quasar surveys (e.g., \cite{reed17}), while more advanced
Bayesian treatment, e.g., with surface density priors \citep{mortlock12}, is not frequently seen in the literature.
For each source, we calculate the Bayesian probability ($P_{\rm Q}^{\rm B}$) of being a high-$z$ quasar rather than a contaminating brown dwarf, 
based on models of SED and surface density as a function of magnitude. 
Galaxy models are not included in the algorithm at present.
The quasar SED models were created by stacking the SDSS spectra of 340 bright quasars at $z \simeq 3$, where the quasar selection is 
fairly complete \citep{richards02, willott05}, and correcting for the effect of IGM absorption \citep{songaila04}.
The quasar surface density was modeled based on the luminosity function taken from \citet{willott10}.
A more detailed description about our Bayesian algorithm may be found in Paper I.

We keep those sources with $P_{\rm Q}^{\rm B} > 0.1$ in the sample of candidates, while removing sources with lower quasar probability.
%Those with lower $P_{\rm Q}^{\rm B}$ are further checked with the SED-fitting algorithm and the simple color-cut criteria described in \S \ref{subsec:auxiliarysample}, 
%and added to the Auxiliary sample if $P_{\rm Q}^{\rm S} > 0.1$ or $P_{\rm Q}^{\rm C} = 1$.
%We calculated $P_{\rm Q}^{\rm S}$ and $P_{\rm Q}^{\rm C}$ for the objects in the Main sample as well and found that they always meet the Auxiliary selection criteria, i.e.,
%the Main sample is a subsample of the Auxiliary sample.
The rather low value of the threshold $P_{\rm Q}^{\rm B} = 0.1$ was chosen to ensure that we would not throw away any possible candidates.
We found that the actual $P_{\rm Q}^{\rm B}$ distribution is bimodal, with only a small fraction falling in  $0.1 < P_{\rm Q}^{\rm B} < 0.9$ (see \S \ref{subsec:efficiency}),
so our results are insensitive to the exact value of this cut.
As we will see below, the bimodal $P_{\rm Q}^{\rm B}$ distribution is populated mostly by high-$z$ quasars at $P_{\rm Q}^{\rm B} \simeq 1$ 
and brown dwarfs at $P_{\rm Q}^{\rm B} \simeq 0$.
This indicates that quasars and brown dwarfs occupy distinct regions of the color space, and HSC (and UKIDSS/VIKING) photometry is
sufficiently accurate to separate these two populations, down to the limiting magnitude of our quasar survey.

Finally, we inspect the images of all the candidates by eye and reject additional problematic sources.
Both stacked and per-visit images are used for this purpose. 
The sources rejected at this stage include those close to bright stars, cosmic rays, and moving objects overlooked in the above automatic procedure.

In the present survey area covering 430 deg$^2$, we had roughly 300,000 red point sources meeting the database query conditions (Equations \ref{eq:query1} and \ref{eq:query2}) 
and undetected in the $g$ and $r$ bands.
The vast majority of them ($\sim$97 \%) were eliminated by the automatic image checking procedure, and the Bayesian algorithm selected 261 final candidates with $P_{\rm Q}^{\rm B} > 0.1$.
Among them, we put highest priority for follow-up spectroscopy on the 60 candidates with reddest colors ($i_{\rm AB} - z_{\rm AB} > 2.0$ or $z_{\rm AB} - y_{\rm AB} > 0.8$; 
see Figure \ref{fig:colordiagram}), relatively
bright magnitudes ($z_{\rm AB} <$ 24 mag), and detection in more than a single band or a single exposure.
%or with detection only in the $z$ band, and obtained xx final candidates after visual image inspection.

\subsection{Recovery of Known Objects \label{subsec:known_quasars}}

%\textcolor{red}{
%In the present HSC survey footprint, there are ten high-$z$ quasars discovered by the previous surveys.
%We have recovered seven of these quasars (VIK $J0839+0015$, IMS $J2204+0012$, CFHQS $J0210-0456$, CFHQS $J0216-0455$, CFHQS $J0227-0605$, VIMOS2911001793, and SDSS $J2228+0110$, 
%following the naming convention of \cite{banados16}), 
%all of which have $P_{\rm Q}^{\rm B} = 1.00$.
%Among the three remaining quasars, the first two (SDSS $J0836+0054$ and VIK $J1148+0056$) have relatively low redshifts ($z$ = 5.81 and 5.84, respectively) and bluer HSC colors ($i - z < 1.5$) than 
%our selection threshold.
%The third quasar (SDSS $J1602+4228$) was dropped from the selection because of its significant $r$-band flux ($r_{\rm AB} = 25.05 \pm 0.06$ mag) in the HSC catalog.
%We checked the HSC images of this last object, and found that it is blended with a nearby faint source at a few arcsec away, which likely contributes to the flux at the quasar position.
%}

In the 430 deg$^2$ surveyed in this paper, there are ten high-$z$ quasars discovered by previous surveys.
We have recovered seven of these quasars (CFHQS $J0210-0456$, CFHQS $J0216-0455$, CFHQS $J0227-0605$, SDSS $J1602+4228$, IMS $J2204+0012$, VIMOS2911001793, and SDSS $J2228+0110$, 
following the naming convention of \cite{banados16}), 
all of which have $P_{\rm Q}^{\rm B} = 1.00$.
On the other hand, two quasars (SDSS $J0836+0054$ and VIK $J1148+0056$) have relatively low redshifts ($z$ = 5.81 and 5.84, respectively) and bluer HSC colors ($i - z < 1.5$) than 
our selection threshold.
The remaining quasar (VIK $J0839+0015$) was dropped from the selection, because it has $r$-band detection ($r_{\rm AB} = 25.16 \pm 0.11$ mag) in %they hav significant ($>3\sigma$) $r$-band fluxes in 
the HSC catalog.
%Indeed we found a faint flux peak at the position of this quasar in the HSC $r$-band image. %, though it is close to a moderately bright star.
This quasar is at $z = 5.84$ and bright also in the $i$ band ($i_{\rm AB} = 22.86 \pm 0.01$ mag), so the $r$-band detection may be real.
Alternatively, the $r$-band flux peak may be an artifact due to the halo around a saturated star, at $\sim$13\arcsec\ away from the quasar.
%We checked the HSC $r$-band image by eye, and found that 
%in all the cases there is a bright or very close nearby source, which likely contributes to the flux at the quasar position.
%Since such flux contamination from surrounding sources reduces the effective survey area, we recently changed our selection algorithm, so that it doesn't remove objects based only on the catalog 
%$g$- or $r$-band detection.
%This new algorithm is being used for our up-to-date selection, whose results will be presented in future papers.

Our quasar selection algorithm was run on all HSC data observed before 2016 April, so the present quasar candidates supersede those in Paper I.
%The above algorithm was run on all the HSC survey fields observed before 2016 April, therefore the present quasar candidates supersede those in our previous work.
Because the HSC data reduction pipeline is improving continuously, and because we've made a minor change of the selection criteria as mentioned above,
the candidate lists in a given observed field may vary from one round of selection to another.
%which also contributes to better candidate selection in the latest algorithm.
In Paper I, we reported spectroscopic identification of nine high-$z$ quasars, six high-$z$ galaxies, and one brown dwarf.
We found that the present selection recovers eight quasars and two galaxies from the above sample.
$J2232+0012$, which we classified as a quasar based on the very high luminosity ($\sim 10^{44.1}$ erg s$^{-1}$) of its narrow Ly$\alpha$ line, was dropped from the candidates, 
because its extendedness ($z_{\rm AB} - z_{\rm CModel, AB} = 0.156$) is slightly above our new threshold (0.150; see above).
The two galaxies $J0857+0142$ and $J0848+0045$ were dropped from the quasar candidates for the same reason.
The three remaining sources, i.e., the galaxies $J0210-0523$ and $J0215-0555$ and the brown dwarf $J0850+0012$, have lower $P_{\rm Q}^{\rm B}$ values than in Paper I, and 
didn't meet the selection criterion ($P_{\rm Q}^{\rm B} > 0.1$).
%We discuss this issue further in \S \ref{subsec:efficiency}.
%\textcolor{red}{Are there new candidates in the Paper I footprint? ... this will be difficult to answer, due to the continuously increasing area and depth of the survey...}

%K1 to K5 are recovered.

%Three objects not recovered due to rmagerr $<$ 0.36 mag, due to nearby stars. This should be revisited!

%Three objects at $z < 5.9$ not recovered because $i - z < 1.5$.

%See check.known.quasars.dat

%The following four objects in paper I had magdiff $>$ 0.15. A2 and A3 are in Paper I, and not recovered in this work???
%quasar     0.193000   J2228+0128   A2
%quasar     0.156000   J2232+0012   A3
%quasar     0.200000   J1416+0015   B11
%quasar     0.483000   J1440-0107   B13

\section{Spectroscopy \label{sec:spectroscopy}}

%\textcolor{red}{
%As we described in the previous section, we have 66 quasar candidates with the highest priority for spectroscopic follow-up.
%Five of these are known quasars discovered by previous surveys, while the other ten candidates were already observed and presented in Paper I.
%We took optical spectra of 48 out of the remaining 51 candidates}
Since the discovery reported in Paper I, 
we took optical spectra of 48 additional unidentified quasar candidates, %described in \S \ref{subsec:selection}, %in the 2016 Spring and Fall semesters, 
using 
the Optical System for Imaging and low-Intermediate-Resolution Integrated Spectroscopy (OSIRIS; \cite{cepa00}) mounted on the Gran Telescopio Canarias (GTC),
and the Faint Object Camera and Spectrograph (FOCAS; \cite{kashikawa02}) mounted on Subaru.
The observations were carried out in the 2016 Spring and Fall semesters.
Roughly the brighter half of the candidates were observed with OSIRIS, while the fainter candidates were observed with FOCAS.
We prioritized observations in such a way that the targets with brighter magnitudes and higher $P_{\rm Q}^{\rm B}$ were observed at the earlier opportunities.
The journal of these discovery observations is presented in Table \ref{tab:obsjournal}.
The details of the observations are described in the following sections.

\begin{longtable}{lrlc}
  \caption{Journal of Discovery Observations}\label{tab:obsjournal}
  \hline\hline
  Target & Exp. Time & Date & Telescope \\ 
\endfirsthead
  \hline
  Target & Exp. Time & Date & Telescope \\ 
\endhead
  \hline
\endfoot
  \hline
\endlastfoot
  \hline
$J1429-0104$ & 300 min & 2016 May 2, 13,  Jun 27, 30 & GTC\\ %60 + 60 + 60 + 60 + 60 min
$J0857+0056$ & 120 min & 2016 Feb 12, 14 & Subaru \\ %60 + 60 min
$J0905+0300$ & 60 min & 2016 Feb 13 & Subaru \\
$J2239+0207$ & 15 min & 2016 Jun 8 & GTC \\
$J0844-0052$ &  60 min & 2016 Apr 9 & GTC \\
$J1208-0200$ &  30 min & 2016 Feb 13 & Subaru \\
$J0217-0208$ &  60 min & 2016 Sep 9 & Subaru \\
$J1425-0015$ &  180 min & 2016 Feb 12, 14, 16 & Subaru \\ %80 + 40 +60 min
$J2201+0155$ & 80 min & 2016 Sep 7 & Subaru \\ %40 + 40 min
$J1423-0018$ &  120 min & 2016 Feb 14, 16 & Subaru \\%60 + 60 min 
$J1440-0107$ &  40 min & 2016 Feb 13 & Subaru \\
$J0235-0532$ &  60 min & 2016 Sep 7 & Subaru \\
$J2228+0152$ & 30 min & 2016 Jun 27 & GTC \\
$J0911+0152$ & 120 min & 2016 Feb 15, 16 & Subaru \\%90 + 30 min
$J1201+0133$ & 120 min & 2016 Feb 12, 14 & Subaru \\%80 + 40 min 
$J1429-0002$ &  60 min & 2016 Feb 12 & Subaru \\
$J0202-0251$ & 45 min & 2016 Aug 10 & GTC\\
$J0206-0255$ & 15 min & 2016 Jul 30 & GTC\\
$J1416+0015$ & 60 min & 2016 Feb 14 & Subaru \\
$J1417+0117$ & 60 min & 2016 Feb 13 & Subaru \\
$J0902+0155$ & 160 min & 2016 Feb 12, 14 & Subaru \\%100 + 60 min
$J0853+0139$ & 30 min & 2016 Feb 15 & Subaru \\
$J1414+0130$ &  60 min & 2016 Apr 13 & GTC\\
$J0903+0211$ & 120 min & 2016 Apr 3, 24 & GTC\\%60 + 60 min 
$J1628+4312$ & 170 min & 2016 Feb 13, 15, 16 & Subaru \\%60 + 30 + 80 min 
$J1211-0118$ &  60 min & 2016 Apr 28 & GTC\\
$J1630+4315$ & 45 min & 2016 Feb 14 & Subaru \\
$J2233+0124$ & 60 min & 2016 Sep 9 & Subaru \\
$J0212-0158$ &  60 min & 2016 Aug 27 & GTC\\
$J0218-0220$ &  60 min & 2016 Sep 7 & Subaru \\
$J0159-0359$ &  60 min & 2016 Sep 9 & Subaru \\
$J2237-0006$ &  100 min & 2016 Sep 9 & Subaru \\
$J1157-0157$ &  45 min & 2016 Apr 9 & GTC \\
$J1443-0214$ &  15 min & 2016 May 5 & GTC \\
$J0210-0451$ &  25 min & 2016 Sep 7 & Subaru\\
$J0211-0414$ & 135 min & 2015 Sep 9, 12 & GTC\\%45 + 45 + 45 min
$J0214-0214$ & 15 min & 2016 Sep 7 & Subaru\\
$J0214-0645$ & 15 min & 2016 Jul 31 & GTC\\
$J0217-0708$ & 135 min & 2015 Sep 12, 14 & GTC\\%45 + 45 + 45 min
$J0226-0403$ & 90 min & 2015 Sep 12 & GTC\\%45 + 45 min
$J0230-0623$ & 135 min & 2015 Sep 8 & GTC\\%45 + 45 + 45 min
$J0234-0604$ & 15 min & 2016 Aug 2 & GTC\\
$J0854-0004$ & 40 min & 2016 Feb 12 & Subaru\\
$J1204-0046$ & 45 min & 2016 Feb 15 & Subaru\\
$J2206+0231$ & 60 min & 2016 Jun 9 & GTC\\
$J2209+0139$ & 20 min & 2016 Sep 7 & Subaru\\
$J2211-0027$ & 25 min & 2016 Sep 7 & Subaru\\ 
$J2237+0239$ & 60 min & 2016 Jun 30 & GTC\\
\end{longtable}

\subsection{GTC/OSIRIS}

GTC is a 10.4-m telescope located at the Observatorio del Roque de los Muchachos in La Palma, Spain.
Our program was awarded 14.4 and 21.5 hours in the 2015B and 2016A semesters, respectively (GTC19-15B and GTC4-16A; Iwasawa et al.).
We used OSIRIS with the R2500I grism and 1\arcsec.0-wide long slit, which provides spectral coverage from $\lambda_{\rm obs}$ = 0.74 to 1.0\ $\mu$m 
with a resolution $R \sim 1500$.
%Our targets were observed in the queue mode on September x,x,x,x under the good weather conditions. seeing 0.7-0.8 sec
%The observation journal is given in Table \ref{tab:osiris}.
%The telluric standard stars were observed in the same nights with the targets.
The observations were carried out in queue mode on dark and gray nights, with mostly photometric (sometimes spectroscopic) sky conditions and the seeing 0\arcsec.6 -- 1\arcsec.2.
The data were reduced using the Image Reduction and Analysis Facility (IRAF).
Bias correction, flat fielding with dome flats, sky subtraction, and 1d extraction were performed in the standard way.
The wavelength was calibrated with reference to sky emission lines.
The flux calibration was tied to white dwarf standard stars (Feige 110, Feige 66, G191-B2B, GD 140, or Ross 640) observed on the same nights.
We corrected for slit losses by scaling the spectra to match the HSC magnitudes in the $z$ and $y$ bands for the $i$- and $z$-band dropouts, respectively.

%One of the quasar candidates was observed in this GTC run and identified to be a quasar at $z = 6.10$.
%Its spectrum is presented in \S \ref{sec:results}.
%In addition, we observed five candidates selected from the older (S14B) version of the HSC-SSP data release.
%These five objects are no longer quasar candidates with the revised photometry in our fiducial (S15A) release, and hence are not included in the sample of
%38 candidates described above.
%Indeed, their OSIRIS spectra show relatively smooth red continua characteristic of brown dwarfs, which are also consistent with the latest HSC magnitudes.
%The detailed analysis of these objects is still underway and will be reported in a future paper.

\subsection{Subaru/FOCAS}

Our program was awarded five nights each in the S16A and S16B semesters (S16A-076 and S16B-071I; Matsuoka et al.) with the Subaru 8.2-m telescope.
The latter program (S16B-071I) has been approved as a Subaru intensive program, for which a total of 20 nights will be allocated during the S16B -- S18A semesters.
We used FOCAS in the multi-object spectrograph mode with the VPH900 grism and SO58 order-sorting filter.
The widths of the slitlets were set to 1\arcsec.0.
This configuration provides spectral coverage from $\lambda_{\rm obs}$ = 0.75 to 1.05\ $\mu$m with a resolution $R \sim 1200$.
All the observations were carried out on grey nights.
A few of these nights were occasionally affected by cirrus and poor seeing ($\lesssim$ 2\arcsec.0), while the weather was fairly good
with seeing 0\arcsec.5 -- 1\arcsec.0 for the rest of the observations.

The data were reduced with IRAF using the dedicated FOCASRED package.
Bias correction, flat fielding with dome flats, sky subtraction, and 1d extraction were performed in the standard way.
The wavelength was calibrated with reference to the sky emission lines.
The flux calibration was tied to white dwarf standard stars (Feige 110 and G191-B2B) observed on the same nights as the targets.
We corrected for slit losses in the same way as in the OSIRIS data reductions.

%We observed 19 targets from the sample of candidates, including the quasar identified with GTC/OSIRIS, and identified 14 new high-$z$ quasars and galaxies
%as well as a brown dwarf.
%Their final spectra are presented in \S \ref{sec:results}.
%One of the $z$-band dropout targets was not found at the HSC position at the time of the spectroscopy, so is most likely a moving object or a transient event
%caught by the HSC $y$-band observations.
%All the $y$-band exposures of this object were taken in a single day, and our inspection of per-visit images did not detect significant day-scale
%motion or flux variation.
%The spectral S/N of the remaining two targets is too low to judge their nature at this moment.

\section{Results and Discussion \label{sec:results}}

\begin{figure*}
 \begin{center}
  \includegraphics[width=16cm]{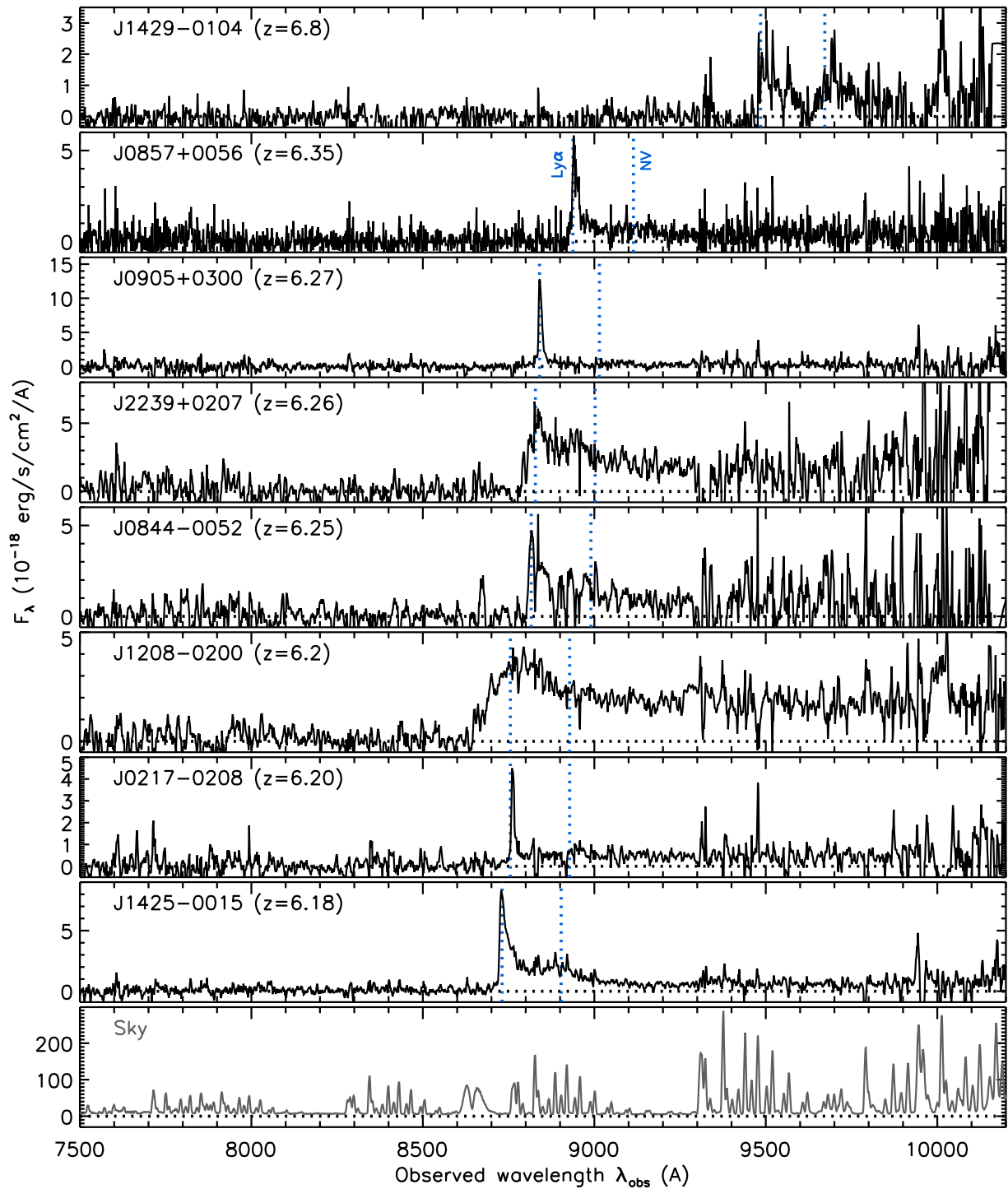} 
 \end{center}
\caption{Reduced spectra of the first set of eight quasars and possible quasars discovered in this work, displayed in decreasing order of redshift.
The object name and the estimated redshift are indicated at the top left corner of each panel.
The blue dotted lines mark the expected positions of the Ly$\alpha$ and N \emissiontype{V} $\lambda$1240 emission lines, given the redshifts.
The spectra were smoothed using inverse-variance weighted means over 3 -- 9 pixels (depending on the S/N), for display purposes.
The bottom panel displays a sky spectrum, as a guide to the expected noise.}
\label{fig:spectra1}
\end{figure*}

\begin{figure*}
 \begin{center}
  \includegraphics[width=16cm]{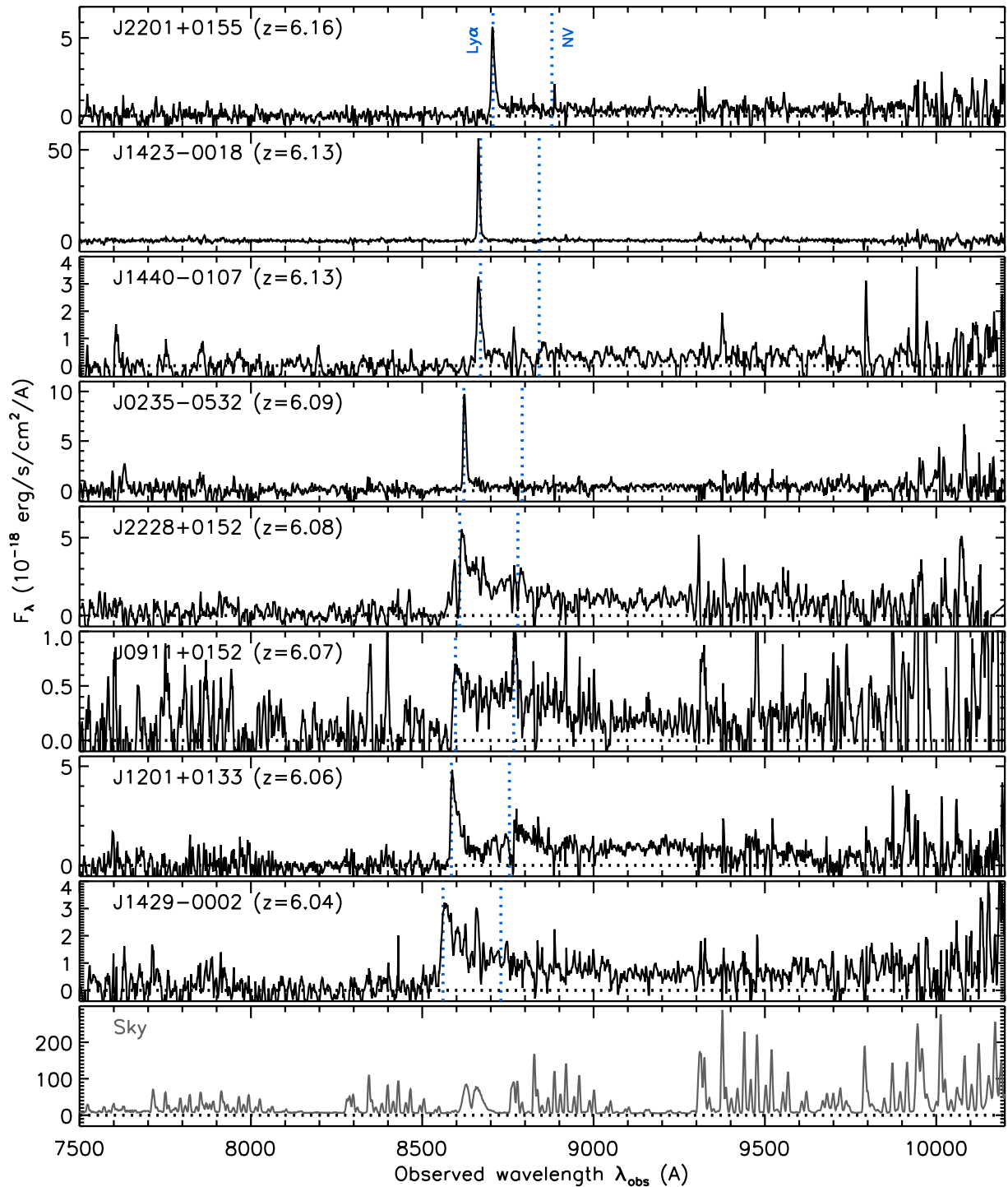} 
 \end{center}
\caption{Same as Figure \ref{fig:spectra1}, but for the second set of eight quasars and possible quasars.}
\label{fig:spectra2}
\end{figure*}

\begin{figure*}
 \begin{center}
  \includegraphics[width=16cm]{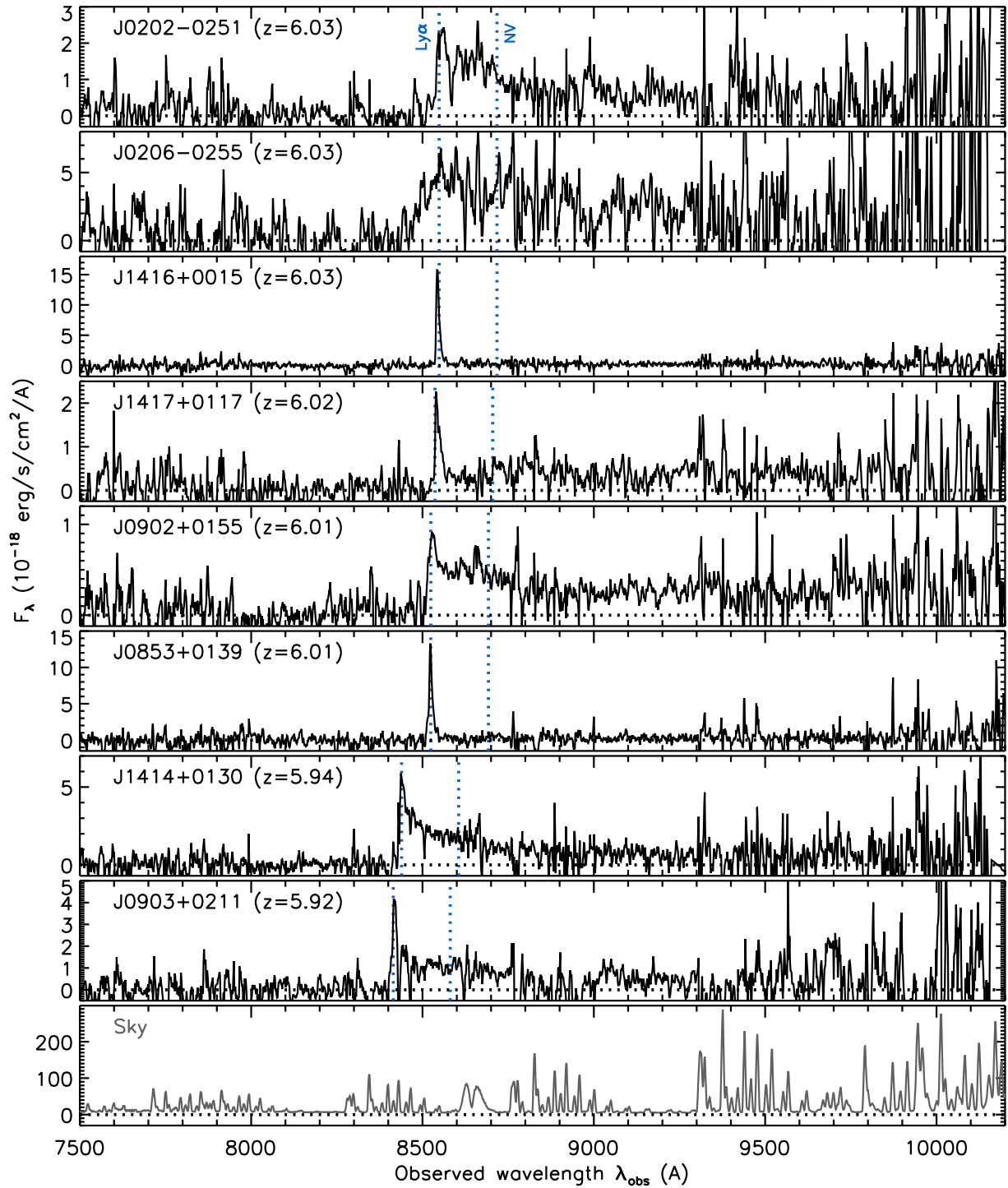} 
 \end{center}
\caption{Same as Figure \ref{fig:spectra1}, but for the last set of eight quasars and possible quasars.}
\label{fig:spectra3}
\end{figure*}

\begin{figure*}
 \begin{center}
  \includegraphics[width=16cm]{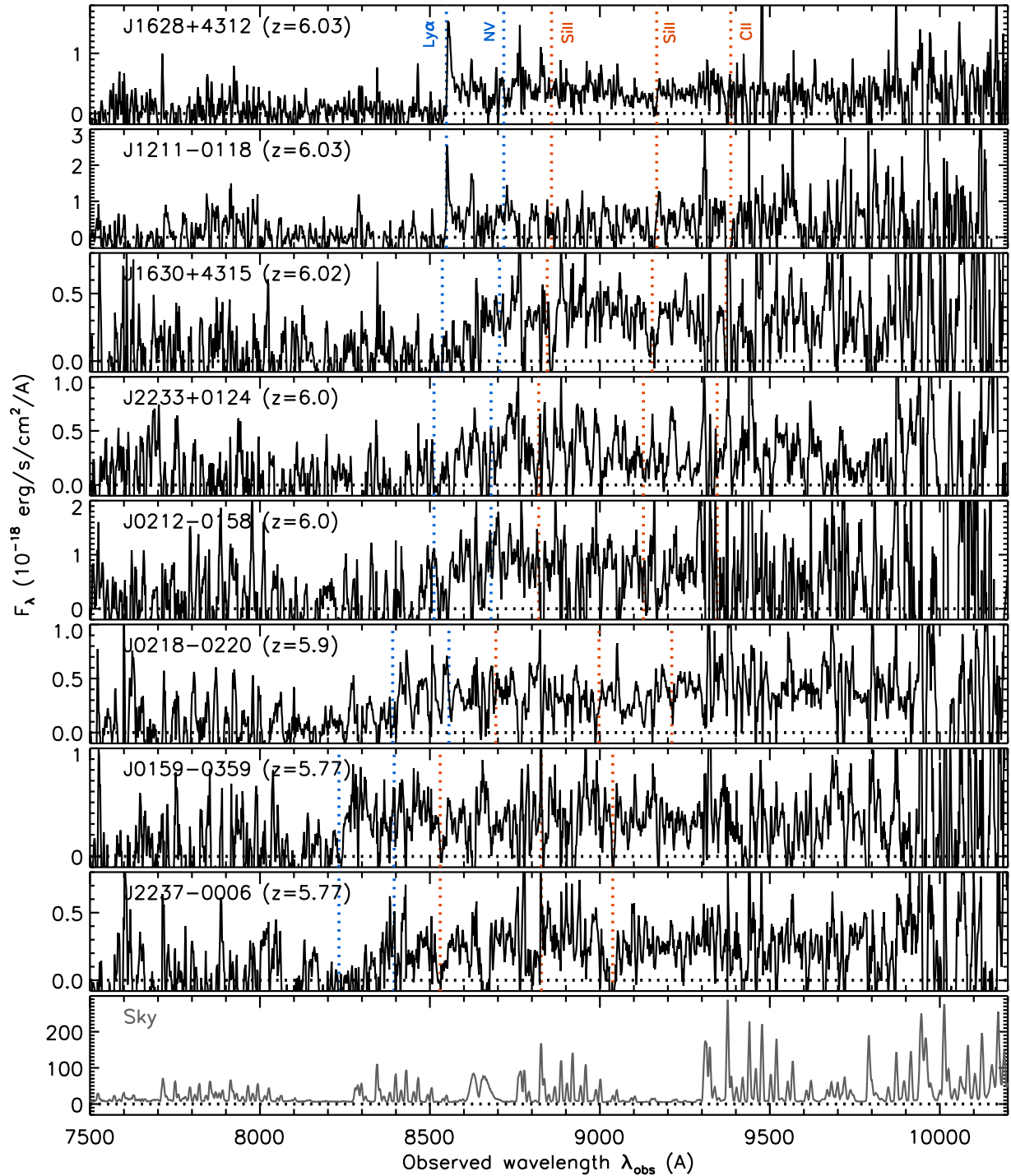} 
 \end{center}
\caption{Same as Figure \ref{fig:spectra1}, but for the eight high-$z$ galaxies.
The expected positions of the interstellar absorption lines of Si \emissiontype{II} $\lambda$1260, Si \emissiontype{II} $\lambda$1304, and C \emissiontype{II} $\lambda$1335 
are marked by the red dotted lines.}
\label{fig:spectra4}
\end{figure*}

\begin{figure*}
 \begin{center}
  \includegraphics[width=16cm]{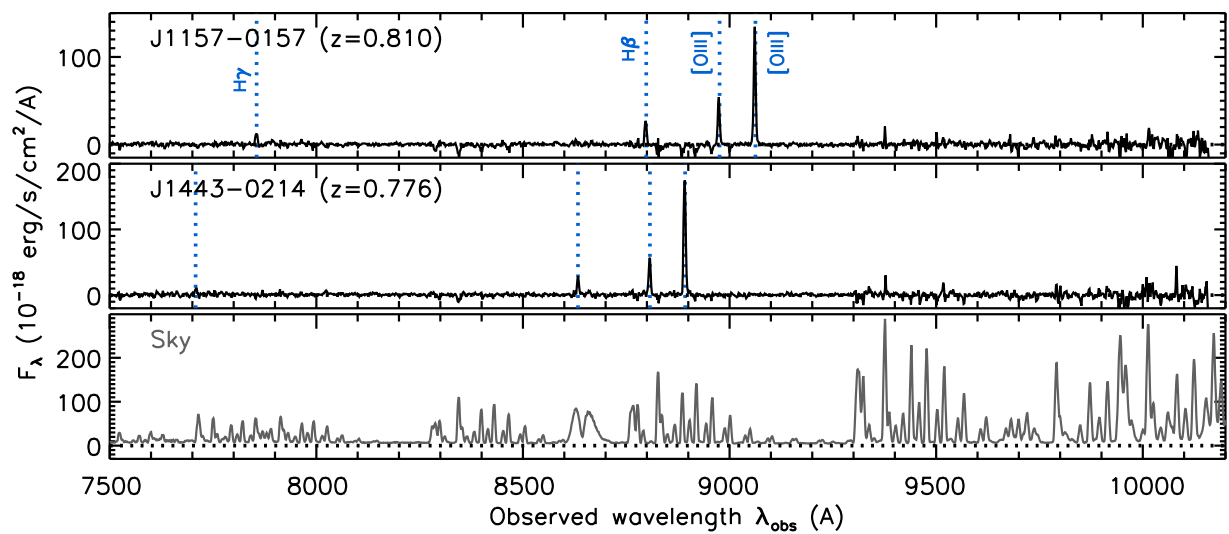} 
 \end{center}
\caption{Same as Figure \ref{fig:spectra1}, but for the two [O \emissiontype{III}] emitters at $z \sim 0.8$.
The expected positions of H$\gamma$, H$\beta$, and two [O \emissiontype{III}] lines ($\lambda$4959 and $\lambda$5007) are marked by the dotted lines.}
\label{fig:spectra5}
\end{figure*}

\begin{figure*}
 \begin{center}
  \includegraphics[width=16cm]{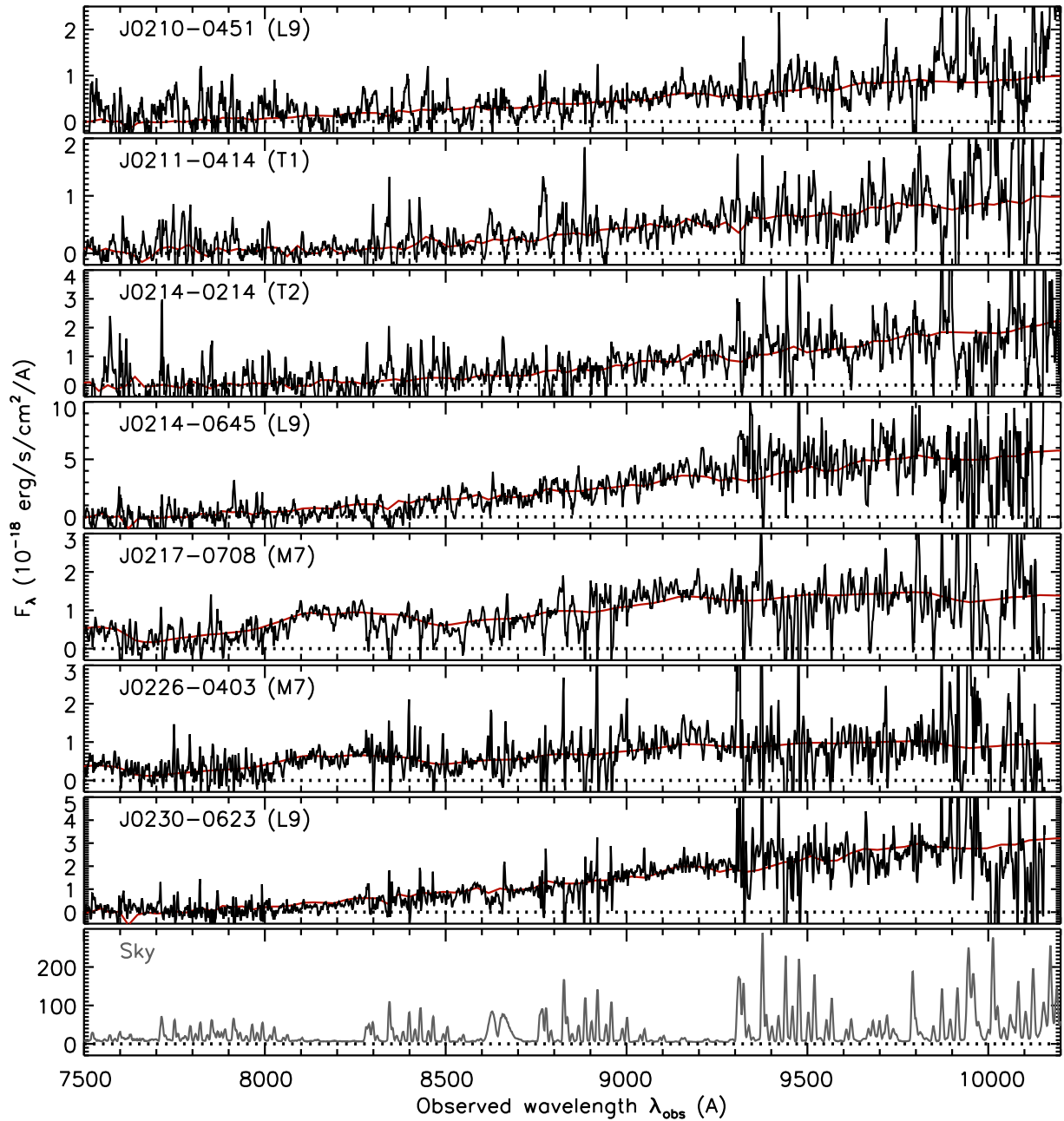} 
 \end{center}
\caption{Same as Figure \ref{fig:spectra1}, but for the first seven brown dwarfs. The red lines represent the best-fit templates, whose spectral types are indicated at the top left corner of each panel.}
\label{fig:spectra6}
\end{figure*}

\begin{figure*}
 \begin{center}
  \includegraphics[width=16cm]{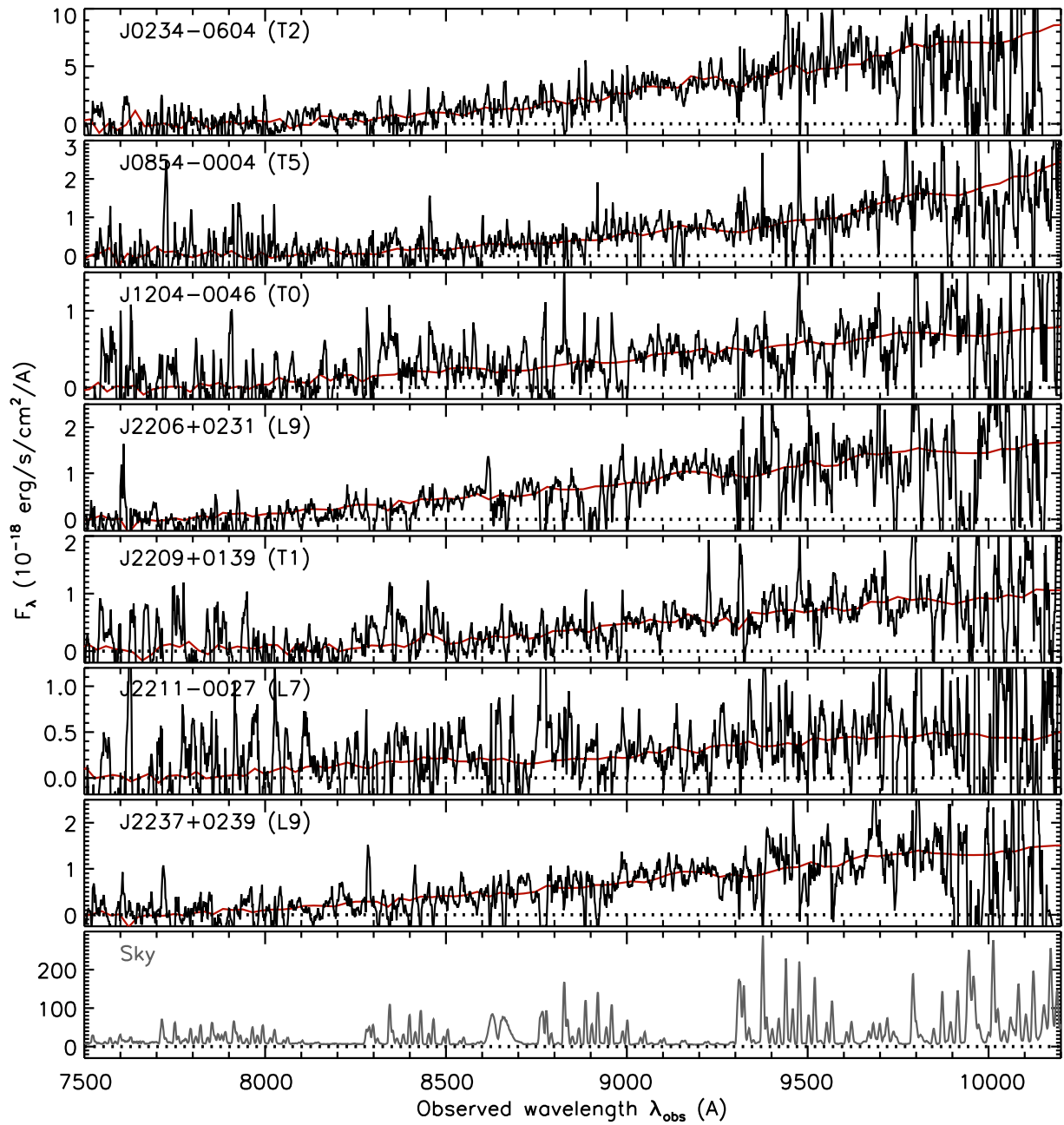} 
 \end{center}
\caption{Same as Figure \ref{fig:spectra6}, but for the remaining seven brown dwarfs.}
\label{fig:spectra7}
\end{figure*}

Figures \ref{fig:spectra1} -- \ref{fig:spectra7} present the reduced spectra of the 48 quasar candidates.
As we describe in detail below, we identified 24 high-$z$ quasars, 8 high-$z$ galaxies, 2 strong [O\emissiontype{III}] emitters at $z \sim 0.8$, and 14 brown dwarfs.
Their photometric properties are summarized in Table \ref{tab:photometry}.
Note that the astrometric accuracy of the HSC-SSP data is estimated to be $\lesssim$ 0\arcsec.1 (root mean square; \cite{aihara17}).
Table \ref{tab:photometry} also lists the updated magnitudes and $P_{\rm Q}^{\rm B}$ of the 16 objects presented in Paper I, measured with the present version of the HSC data reduction pipeline.
%for the 16 objects presented in Paper I (9 high-$z$ quasars, 6 high-$z$ galaxies, and 1 brown dwarf).
Note that some of the objects in this table do not meet our latest quasar selection criteria, due either to the improvement 
of the HSC photometry or to our minor changes in the selection criteria (see \S \ref{subsec:selection}).
We clarify this point below whenever necessary.
We have now spectroscopically identified 64 HSC sources in Paper I and this work, which include 33 high-$z$ quasars, 14 high-$z$ galaxies, 2 [O\emissiontype{III}] emitters,
and 15 brown dwarfs.
Six of these sources are detected in the $J$, $H$, and/or $K$ band, as summarized in Table \ref{tab:nir_photometry}.
%The $JHK$ magnitudes of the objects detected in UKIDSS or VIKING are listed in , which is used for discussion in \S \ref{subsec:efficiency}.

\begin{longtable}{lcccccc}
  \caption{Photometric properties}\label{tab:photometry}
  \hline\hline              
  Name & R.A. & Decl. & $i_{\rm AB}$ (mag)  & $z_{\rm AB}$ (mag) & $y_{\rm AB}$ (mag) & $P_{\rm Q}^{\rm B}$ \\ 
\endfirsthead
  \hline
  Name & R.A. & Decl. & $i_{\rm AB}$ (mag)  & $z_{\rm AB}$ (mag) & $y_{\rm AB}$ (mag) & $P_{\rm Q}^{\rm B}$ \\ 
\endhead
  \hline
\endfoot
  \hline 
  \multicolumn{7}{l}{\hbox to 0pt{\parbox{170mm}{\footnotesize
    \footnotemark[] Note --- Coordinates are at J2000.0. Magnitude upper limits are placed at $5\sigma$ significance.
     Names with * denote the objects taken from Paper I.
 }}}
\endlastfoot
  \hline
  \multicolumn{7}{c}{High-$z$ Quasars} \\ 
  \hline
  $J1429-0104$ & 14:29:03.08 & $-$01:04:43.4 &         $>$26.61 &          $>$25.25 & 23.73 $\pm$ 0.09 &         0.86     \\ % a24 (ju=0.000 $\pm$ 9.999)
  $J0857+0056$ & 08:57:38.53 & $+$00:56:12.7 & 27.43 $\pm$ 0.97 & 24.08 $\pm$ 0.05 & 24.14 $\pm$ 0.14 &         1.00     \\ % a12 (ju=0.000 $\pm$ 9.999)
  $J0905+0300$ & 09:05:44.65 & $+$03:00:58.8 & 26.94 $\pm$ 0.29 & 24.16 $\pm$ 0.06 & 24.18 $\pm$ 0.14 &         1.00     \\ % a14 (ju=0.000 $\pm$ 9.999)
  $J2239+0207$ & 22:39:47.47 & $+$02:07:47.5 & 25.60 $\pm$ 0.09 & 22.40 $\pm$ 0.01 & 22.33 $\pm$ 0.03 &         1.00     \\ % s18 (ju=0.000 $\pm$ 9.999)
  $J0844-0052$ & 08:44:31.60 & $-$00:52:54.6 &          $>$25.65 & 23.18 $\pm$ 0.03 & 23.12 $\pm$ 0.08 &         1.00     \\ % a22 (ju=0.000 $\pm$ 9.999)
  $J1208-0200$ & 12:08:59.23 & $-$02:00:34.8 & 24.65 $\pm$ 0.08 & 22.13 $\pm$ 0.02 & 22.05 $\pm$ 0.03 &         1.00     \\ % s10 (ju=0.000 $\pm$ 9.999)
  $J0217-0208$ & 02:17:21.59 & $-$02:08:52.6 &          $>$25.88 & 23.88 $\pm$ 0.04 & 23.50 $\pm$ 0.08 &         1.00     \\ % a31 (ju=0.000 $\pm$ 9.999)
  $J1425-0015$ & 14:25:17.72 & $-$00:15:40.9 & 26.30 $\pm$ 0.14 & 22.82 $\pm$ 0.02 & 23.37 $\pm$ 0.05 &         1.00      \\ % s5 (ju=0.000 $\pm$ 9.999)
  $J2201+0155$ & 22:01:32.07 & $+$01:55:29.0 & 27.49 $\pm$ 0.89 & 23.98 $\pm$ 0.08 & 24.29 $\pm$ 0.19 &         1.00     \\ % a27 (ju=0.000 $\pm$ 9.999)
  $J1423-0018$ & 14:23:31.71 & $-$00:18:09.1 &          $>$26.64 & 24.18 $\pm$ 0.06 & 24.79 $\pm$ 0.23 &         1.00     \\ % b12 (ju=0.000 $\pm$ 9.999)
  $J1440-0107$ & 14:40:01.30 & $-$01:07:02.2 & 26.98 $\pm$ 0.41 & 24.14 $\pm$ 0.06 & 24.01 $\pm$ 0.10 &         1.00     \\ % b13 (ju=0.000 $\pm$ 9.999)
  $J0235-0532$ & 02:35:42.42 & $-$05:32:41.6 & 27.24 $\pm$ 0.57 & 23.77 $\pm$ 0.06 & 23.96 $\pm$ 0.15 &         1.00     \\ % a26 (ju=0.000 $\pm$ 9.999)
  $J2228+0152$ & 22:28:47.71 & $+$01:52:40.5 & 25.57 $\pm$ 0.09 & 22.67 $\pm$ 0.02 & 22.91 $\pm$ 0.04 &         1.00     \\ % a23 (ju=0.000 $\pm$ 9.999)
  $J0911+0152$ & 09:11:14.27 & $+$01:52:19.4 & 27.81 $\pm$ 0.84 & 24.22 $\pm$ 0.08 & 24.35 $\pm$ 0.13 &         1.00     \\ % b23 (ju=0.000 $\pm$ 9.999)
  $J1201+0133$ & 12:01:03.02 & $+$01:33:56.4 &          $>$25.53 & 22.94 $\pm$ 0.03 & 23.31 $\pm$ 0.09 &         1.00      \\ % s9 (ju=0.000 $\pm$ 9.999)
  $J1429-0002$ & 14:29:20.22 & $-$00:02:07.4 & 26.00 $\pm$ 0.12 & 22.93 $\pm$ 0.02 & 23.27 $\pm$ 0.05 &         1.00      \\ % s6 (ju=0.000 $\pm$ 9.999)
  $J0202-0251$ & 02:02:58.21 & $-$02:51:53.6 & 26.39 $\pm$ 0.30 & 23.06 $\pm$ 0.03 & 23.18 $\pm$ 0.04 &         1.00     \\ % s11 (ju=0.000 $\pm$ 9.999)
  $J0206-0255$ & 02:06:11.20 & $-$02:55:37.8 & 24.84 $\pm$ 0.07 & 21.70 $\pm$ 0.01 & 21.88 $\pm$ 0.02 &         1.00     \\ % s16 (ju=0.000 $\pm$ 9.999)
  $J1416+0015$ & 14:16:12.71 & $+$00:15:46.2 & 27.18 $\pm$ 0.38 & 24.15 $\pm$ 0.06 & 23.76 $\pm$ 0.10 &         1.00     \\ % b11 (ju=0.000 $\pm$ 9.999)
  $J1417+0117$ & 14:17:28.67 & $+$01:17:12.4 & 26.56 $\pm$ 0.28 & 23.90 $\pm$ 0.06 & 23.71 $\pm$ 0.12 &         1.00      \\ % a7 (ju=0.000 $\pm$ 9.999)
  $J0902+0155$ & 09:02:54.87 & $+$01:55:10.9 & 26.75 $\pm$ 0.36 & 24.03 $\pm$ 0.04 & 24.32 $\pm$ 0.12 &         1.00     \\ % a13 (ju=0.000 $\pm$ 9.999)
  $J0853+0139$ & 08:53:48.84 & $+$01:39:11.0 & 26.88 $\pm$ 0.47 & 24.23 $\pm$ 0.06 & 24.12 $\pm$ 0.11 &         1.00     \\ % b17 (ju=0.000 $\pm$ 9.999)
  $J1414+0130$ & 14:14:39.54 & $+$01:30:36.5 & 25.26 $\pm$ 0.12 & 22.90 $\pm$ 0.03 & 23.16 $\pm$ 0.06 &         1.00     \\ % s14 (ju=0.000 $\pm$ 9.999)
  $J0903+0211$ & 09:03:14.68 & $+$02:11:28.3 & 25.30 $\pm$ 0.07 & 23.61 $\pm$ 0.03 & 23.71 $\pm$ 0.07 &         1.00      \\ % c5 (ju=0.000 $\pm$ 9.999)
$J1205-0000^*$ & 12:05:05.09 & $-$00:00:27.9 &          $>$26.61 &          $>$25.92         & 22.61 $\pm$ 0.03 &         1.00      \\ % s2 (ju=0.000 $\pm$ 9.999)
$J2236+0032^*$ & 22:36:44.58 & $+$00:32:56.8 &          $>$26.94 & 23.93 $\pm$ 0.04 & 23.19 $\pm$ 0.05 &         1.00      \\ % a4 (ju=0.000 $\pm$ 9.999)
$J0859+0022^*$ & 08:59:07.19 & $+$00:22:55.9 & 27.55 $\pm$ 0.84 & 22.77 $\pm$ 0.01 & 23.62 $\pm$ 0.07 &         1.00      \\ % c1 (ju=0.000 $\pm$ 9.999)
$J1152+0055^*$ & 11:52:21.27 & $+$00:55:36.6 & 25.43 $\pm$ 0.09 & 21.77 $\pm$ 0.01 & 21.57 $\pm$ 0.02 &         1.00      \\ % s4 (ju=0.000 $\pm$ 9.999)
$J2232+0012^*$ & 22:32:12.03 & $+$00:12:38.4 & 27.58 $\pm$ 0.47 & 23.84 $\pm$ 0.05 & 24.23 $\pm$ 0.13 &         1.00      \\ % a3 (ju=0.000 $\pm$ 9.999)
$J2216-0016^*$ & 22:16:44.47 & $-$00:16:50.0 & 25.97 $\pm$ 0.15 & 22.76 $\pm$ 0.03 & 22.94 $\pm$ 0.04 &         1.00      \\ % s1 (ju=0.000 $\pm$ 9.999)
$J2228+0128^*$ & 22:28:27.83 & $+$01:28:09.4 & 27.38 $\pm$ 0.40 & 24.05 $\pm$ 0.05 & 24.50 $\pm$ 0.15 &         1.00      \\ % a2 (ju=0.000 $\pm$ 9.999)
$J1207-0005^*$ & 12:07:54.14 & $-$00:05:53.2 & 26.34 $\pm$ 0.16 & 23.98 $\pm$ 0.04 & 23.83 $\pm$ 0.09 &         1.00      \\ % a6 (ju=0.000 $\pm$ 9.999)
$J1202-0057^*$ & 12:02:46.37 & $-$00:57:01.6 & 26.11 $\pm$ 0.14 & 23.77 $\pm$ 0.03 & 23.77 $\pm$ 0.10 &         1.00      \\ % a5 (ju=0.000 $\pm$ 9.999)
  \hline
  \multicolumn{7}{c}{High-$z$ Galaxies} \\ 
  \hline
  $J1628+4312$ & 16:28:33.02 & $+$43:12:10.6 & 27.52 $\pm$ 0.47 & 23.98 $\pm$ 0.06 & 23.99 $\pm$ 0.13 &         1.00      \\ % a9 (ju=0.000 $\pm$ 9.999)
  $J1211-0118$ & 12:11:37.10 & $-$01:18:16.4 &          $>$26.27 & 23.99 $\pm$ 0.07 & 23.97 $\pm$ 0.10 &         1.00     \\ % a25 (ju=0.000 $\pm$ 9.999)
  $J1630+4315$ & 16:30:26.36 & $+$43:15:58.6 & 26.94 $\pm$ 0.23 & 24.15 $\pm$ 0.07 & 24.08 $\pm$ 0.13 &         1.00     \\ % b15 (ju=0.000 $\pm$ 9.999)
  $J2233+0124$ & 22:33:39.34 & $+$01:24:32.4 & 27.07 $\pm$ 0.29 & 24.20 $\pm$ 0.06 & 24.48 $\pm$ 0.14 &         1.00     \\ % b49 (ju=0.000 $\pm$ 9.999)
  $J0212-0158$ & 02:12:44.75 & $-$01:58:24.6 & 25.53 $\pm$ 0.17 & 23.23 $\pm$ 0.03 & 22.96 $\pm$ 0.06 &         1.00     \\ % s12 (ju=0.000 $\pm$ 9.999)
  $J0218-0220$ & 02:18:03.42 & $-$02:20:29.7 & 26.40 $\pm$ 0.30 & 23.94 $\pm$ 0.04 & 23.56 $\pm$ 0.08 &         1.00     \\ % b24 (ju=0.000 $\pm$ 9.999)
  $J0159-0359$ & 01:59:49.36 & $-$03:59:45.2 & 26.24 $\pm$ 0.24 & 23.96 $\pm$ 0.06 & 24.14 $\pm$ 0.16 &         1.00     \\ % a29 (ju=0.000 $\pm$ 9.999)
  $J2237-0006$ & 22:37:13.51 & $-$00:06:12.7 & 27.62 $\pm$ 0.55 & 24.27 $\pm$ 0.05 & 24.12 $\pm$ 0.11 &         1.00     \\ % b50 (ju=0.000 $\pm$ 9.999)
$J0219-0416^*$ & 02:19:29.41 & $-$04:16:45.9 &          $>$26.49 & 24.27 $\pm$ 0.07 & 24.01 $\pm$ 0.11 &         1.00      \\ % b4 (ju=0.000 $\pm$ 9.999)
$J0210-0523^*$ & 02:10:33.82 & $-$05:23:04.3 & 25.79 $\pm$ 0.17 & 23.73 $\pm$ 0.06 & 23.38 $\pm$ 0.10 &         0.09      \\ % b1 (ju=0.000 $\pm$ 9.999)
$J0857+0142^*$ & 08:57:23.95 & $+$01:42:54.6 & 26.14 $\pm$ 0.25 & 24.12 $\pm$ 0.05 & 23.73 $\pm$ 0.08 &         0.00      \\ % b6 (ju=0.000 $\pm$ 9.999)
$J0210-0559^*$ & 02:10:41.28 & $-$05:59:17.9 & 26.48 $\pm$ 0.27 & 24.24 $\pm$ 0.07 & 24.10 $\pm$ 0.16 &         0.97      \\ % b2 (ju=0.000 $\pm$ 9.999)
$J0848+0045^*$ & 08:48:18.33 & $+$00:45:09.5 & 26.22 $\pm$ 0.22 & 23.82 $\pm$ 0.06 & 23.90 $\pm$ 0.09 &         1.00      \\ % a1 (ju=0.000 $\pm$ 9.999)
$J0215-0555^*$ & 02:15:45.20 & $-$05:55:29.0 & 25.96 $\pm$ 0.16 & 23.97 $\pm$ 0.05 & 23.60 $\pm$ 0.10 &         0.06      \\ % b3 (ju=0.000 $\pm$ 9.999)
  \hline
  \multicolumn{7}{c}{[O \emissiontype{III}] Emitters} \\ 
  \hline
  $J1157-0157$ & 11:57:51.82 & $-$01:57:09.9 & 24.76 $\pm$ 0.09 & 22.94 $\pm$ 0.04 & 24.65 $\pm$ 0.34 &         1.00     \\ % b32 (ju=0.000 $\pm$ 9.999)
  $J1443-0214$ & 14:43:58.26 & $-$02:14:47.3 & 23.90 $\pm$ 0.04 & 22.39 $\pm$ 0.02 & 23.97 $\pm$ 0.18 &         1.00     \\ % b33 (ju=0.000 $\pm$ 9.999)
  \hline
  \multicolumn{7}{c}{Brown Dwarfs} \\ 
  \hline
  $J0210-0451$ & 02:10:47.24 & $-$04:51:03.9 &          $>$25.92 & 23.73 $\pm$ 0.06 & 22.78 $\pm$ 0.05 &         0.14     \\ % a19 (ju=0.000 $\pm$ 9.999)
  $J0211-0414$ & 02:11:25.26 & $-$04:14:03.5 & 26.83 $\pm$ 0.37 & 23.96 $\pm$ 0.06 & 22.43 $\pm$ 0.03 &         0.00      \\ % g1 (ju=0.000 $\pm$ 9.999)
  $J0214-0214$ & 02:14:25.22 & $-$02:14:59.0 & 26.58 $\pm$ 0.41 & 23.32 $\pm$ 0.03 & 22.28 $\pm$ 0.03 &         0.12     \\ % a15 (ju=0.000 $\pm$ 9.999)
  $J0214-0645$ & 02:14:32.59 & $-$06:45:22.3 & 24.80 $\pm$ 0.10 & 21.84 $\pm$ 0.02 & 20.82 $\pm$ 0.01 &         1.00     \\ % s13 (ju=0.000 $\pm$ 9.999)
  $J0217-0708$ & 02:17:29.47 & $-$07:08:19.6 & 23.94 $\pm$ 0.07 & 22.78 $\pm$ 0.07 & 22.14 $\pm$ 0.08 &         0.00      \\ % g2 (ju=0.000 $\pm$ 9.999)
  $J0226-0403$ & 02:26:18.44 & $-$04:03:06.7 & 24.52 $\pm$ 0.04 & 23.19 $\pm$ 0.04 & 22.58 $\pm$ 0.04 &         0.00      \\ % g4 (ju=0.000 $\pm$ 9.999)
  $J0230-0623$ & 02:30:46.80 & $-$06:23:56.7 & 25.39 $\pm$ 0.21 & 22.50 $\pm$ 0.03 & 21.53 $\pm$ 0.03 &         0.14      \\ % g5 (ju=0.000 $\pm$ 9.999)
  $J0234-0604$ & 02:34:30.10 & $-$06:04:56.5 & 24.79 $\pm$ 0.10 & 21.99 $\pm$ 0.01 & 20.91 $\pm$ 0.01 &         0.06     \\ % s17 (ju=0.000 $\pm$ 9.999)
  $J0854-0004$ & 08:54:10.91 & $-$00:04:54.7 & 27.02 $\pm$ 0.42 & 23.54 $\pm$ 0.03 & 22.62 $\pm$ 0.03 &         0.00      \\ % a8 (ju=0.000 $\pm$ 9.999)
  $J1204-0046$ & 12:04:49.68 & $-$00:46:17.2 & 25.97 $\pm$ 0.12 & 23.94 $\pm$ 0.05 & 23.20 $\pm$ 0.06 &         0.00     \\ % b21 (ju=0.000 $\pm$ 9.999)
  $J2206+0231$ & 22:06:14.53 & $+$02:31:38.9 & 25.78 $\pm$ 0.18 & 23.25 $\pm$ 0.04 & 22.57 $\pm$ 0.05 &         0.29     \\ % a16 (ju=0.000 $\pm$ 9.999)
  $J2209+0139$ & 22:09:06.22 & $+$01:39:57.0 & 26.75 $\pm$ 0.24 & 23.72 $\pm$ 0.03 & 22.69 $\pm$ 0.03 &         0.12     \\ % a21 (ju=0.000 $\pm$ 9.999)
  $J2211-0027$ & 22:11:55.16 & $-$00:27:36.1 &          $>$26.07 & 24.11 $\pm$ 0.08 & 23.37 $\pm$ 0.11 &         0.19     \\ % b48 (ju=0.000 $\pm$ 9.999)
  $J2237+0239$ & 22:37:12.37 & $+$02:39:22.6 & 26.26 $\pm$ 0.18 & 23.30 $\pm$ 0.04 & 22.46 $\pm$ 0.04 &         0.15     \\ % a17 (ju=0.000 $\pm$ 9.999)
$J0850+0012^*$ & 08:50:02.63 & $+$00:12:10.0 & 27.72 $\pm$ 0.84 & 24.04 $\pm$ 0.06 & 23.22 $\pm$ 0.05 &         0.05      \\ % b5 (ju=0.000 $\pm$ 9.999)
 \end{longtable}

\begin{table*}
  \tbl{$JHK$ magnitudes of the objects detected in UKIDSS or VIKING.}{%
  \begin{tabular}{lccccccl}
      \hline
                 & \multicolumn{3}{c}{UKIDSS}  & \multicolumn{3}{c}{VIKING}  & \\
      Name & $J_{\rm AB}$ (mag) & $H_{\rm AB}$ (mag) & $K_{\rm AB}$ (mag) & $J_{\rm AB}$ (mag) & $H_{\rm AB}$ (mag) & $K_{\rm AB}$ (mag) & Comment\\ 
      \hline
%     $J1205-0000^*$   &  ... &    ...                            &    ...                             &  21.951 $\pm$ 0.213 & 21.487 $\pm$ 0.344 & 20.733 $\pm$ 0.182  & quasar\\%S2   1.000    -      -      -    1.000  1.000  1.000     quasar 
%     $J1152+0055^*$   &  ...  &   ...                            &    ...                             &  21.664 $\pm$ 0.223 & $>$ 21.126               & $>$ 21.185                 & quasar\\%S4   1.000    -      -      -    1.000  1.000  1.000     quasar   
%     $J0854-0004$   &  ...  &   ...                             &   ...                             & 21.268 $\pm$ 0.108  & 21.163 $\pm$ 0.279 & $>$ 21.182                 & dwarf\\%A8   0.605    -      -      -    0.000  0.000  0.000      dwarf   
%    $J1204-0046$  &  ...  &   ...                            &    ...                             &  $>$ 21.773              &  20.931 $\pm$ 0.206 & 20.883 $\pm$ 0.214 &  dwarf\\%B21   0.000    -      -      -    0.000  0.000  0.000      dwarf   
%    $J2206+0231$  &  ...  & 20.242 $\pm$ 0.218 &    ...                             &   ...                             &   ...                            &  ...                             & dwarf\\%A16   0.294    -      -      -      -      -      -        dwarf 
%    $J0850+0012^*$   &   ...  &   ...                            &  20.550 $\pm$ 0.271 &  $>$ 21.762                &   $>$ 21.120             & $>$ 21.181              & dwarf\\%B5   0.019    -      -      -    0.026  0.028  0.047      dwarf   
     $J1205-0000^*$   &  ... &    ...                            &    ...                             &  21.95 $\pm$ 0.21 & 21.49 $\pm$ 0.34 & 20.73 $\pm$ 0.18  & high-$z$ quasar\\%S2   1.000    -      -      -    1.000  1.000  1.000     quasar 
     $J1152+0055^*$   &  ...  &   ...                            &    ...                             &  21.66 $\pm$ 0.22 & ...                             & ...                               & high-$z$ quasar\\%S4   1.000    -      -      -    1.000  1.000  1.000     quasar   
     $J0854-0004$   &  ...  &   ...                             &   ...                             & 21.27 $\pm$ 0.11  & 21.16 $\pm$ 0.28 & ...                                 & brown dwarf\\%A8   0.605    -      -      -    0.000  0.000  0.000      dwarf   
    $J1204-0046$  &  ...  &   ...                            &    ...                             &  ...                                &  20.93 $\pm$ 0.21 & 20.88 $\pm$ 0.21 &  brown dwarf\\%B21   0.000    -      -      -    0.000  0.000  0.000      dwarf   
    $J2206+0231$  &  ...  & 20.24 $\pm$ 0.22 &    ...                             &   ...                             &   ...                            &  ...                             & brown dwarf\\%A16   0.294    -      -      -      -      -      -        dwarf 
    $J0850+0012^*$   &   ...  &   ...                            &  20.55 $\pm$ 0.27 &  ...                &   ...             & ...              & brown dwarf\\%B5   0.019    -      -      -    0.026  0.028  0.047      dwarf   
      \hline
    \end{tabular}}\label{tab:nir_photometry}
\begin{tabnote}
$^*$These sources are taken from Paper I.
\end{tabnote}
\end{table*}

\begin{longtable}{lcccccc}
  \caption{Spectroscopic properties}\label{tab:spectroscopy}
  \hline\hline              
  Name & Redshift$^\dagger$ & $M_{1450}$ & Line & EW$_{\rm rest}$ (\AA) & FWHM (km s$^{-1}$) & log $L$ (erg s$^{-1}$) \\ 
\endfirsthead
  \hline
  Name & Redshift$^\dagger$ & $M_{1450}$ & Line & EW$_{\rm rest}$ (\AA) & FWHM (km s$^{-1}$) & log $L$ (erg s$^{-1}$) \\ 
\endhead
  \hline
\endfoot
  \hline
 \multicolumn{7}{l}{\hbox to 0pt{\parbox{180mm}{\footnotesize
    \footnotemark[$*$] These sources are taken from Paper I.
    \footnotemark[$\dagger$] Recorded to two significant figures when the position of Ly$\alpha$ emission or interstellar absorption is unambiguous. 
 }}}
\endlastfoot
  \hline
  \multicolumn{7}{c}{High-$z$ Quasars} \\ 
  \hline
  $J1429-0104$ &   6.8 &   $-23.00 \pm 0.26$             & Ly$\alpha$ &     72 $\pm$ 20 &     1400 $\pm$ 100 &    43.95 $\pm$ 0.06 \\ % a24
  $J0857+0056$ &  6.35 &   $-23.01 \pm 0.07$           & Ly$\alpha$ &     57 $\pm$ 5   &     620 $\pm$ 90     &    43.85 $\pm$ 0.02 \\ % a12
  $J0905+0300$ &  6.27 &   $-22.55 \pm 0.11$           & Ly$\alpha$ &     82 $\pm$ 6   &     250 $\pm$ 40     &    43.89 $\pm$ 0.02 \\ % a14
  $J2239+0207$ &  6.26 &   $-24.69 \pm 0.04$           & Ly$\alpha$ &     32 $\pm$ 3 &     5800 $\pm$ 2700 &    44.27 $\pm$ 0.03 \\ % s18
  $J0844-0052$ &  6.25 &   $-23.74 \pm 0.23$           & Ly$\alpha$ &     34 $\pm$ 13 &     1800 $\pm$ 900  &    43.91 $\pm$ 0.14 \\ % a22
  $J1208-0200$ &   6.2 &   $-24.73 \pm 0.02$            & Ly$\alpha$ &     15 $\pm$ 1  &     5500 $\pm$ 1800 &    43.97 $\pm$ 0.04 \\ % s10
  $J0217-0208$ &  6.20 &   $-23.19 \pm 0.04$           & Ly$\alpha$ &     15 $\pm$ 1  &             $<$ 230        &    43.33 $\pm$ 0.04 \\ % a31
  $J1425-0015$ &  6.18 &   $-23.44 \pm 0.02$           & Ly$\alpha$ &     116 $\pm$ 3 &     1400 $\pm$ 400  &    44.32 $\pm$ 0.01 \\  % s5
  $J2201+0155$ &  6.16 &   $-22.97 \pm 0.04$           & Ly$\alpha$ &     24 $\pm$ 1 &     320 $\pm$ 70       &    43.46 $\pm$ 0.01 \\ % a27
  $J1423-0018$ &  6.13 &   $-21.88 \pm 0.20$           & Ly$\alpha$ &     370 $\pm$ 30 &             $<$ 230     &    44.30 $\pm$ 0.01 \\ % b12
  $J1440-0107$ &  6.13 &   $-22.59 \pm 0.10$           & Ly$\alpha$ &     21 $\pm$ 2 &     440 $\pm$ 260      &    43.27 $\pm$ 0.03 \\ % b13
  $J0235-0532$ &  6.09 &   $-23.01 \pm 0.05$           & Ly$\alpha$ &     41 $\pm$ 2 &     270 $\pm$ 30        &    43.68 $\pm$ 0.02 \\ % a26
  $J2228+0152$ &  6.08 &   $-24.00 \pm 0.04$           & Ly$\alpha$ &     39 $\pm$ 3 &     3000 $\pm$ 200   &    44.07 $\pm$ 0.03 \\ % a23
  $J0911+0152$ &  6.07 &   $-22.09 \pm 0.07$           & Ly$\alpha$ &     77 $\pm$ 8 &     6500 $\pm$ 4200 &    43.60 $\pm$ 0.04 \\ % b23
  $J1201+0133$ &  6.06 &   $-23.85 \pm 0.02$           & Ly$\alpha$ &     17 $\pm$ 1 &     1300 $\pm$ 600   &    43.67 $\pm$ 0.04 \\  % s9
  $J1429-0002$ &  6.04 &   $-23.42 \pm 0.04$           & Ly$\alpha$ &     50 $\pm$ 3 &     2900 $\pm$ 600    &   43.95 $\pm$ 0.02 \\  % s6
  $J0202-0251$ &  6.03 &   $-23.39 \pm 0.07$           & Ly$\alpha$ &     44 $\pm$ 5 &     5600 $\pm$ 800    &   43.88 $\pm$ 0.04 \\ % s11
  $J0206-0255$ &  6.03 &   $-24.91 \pm 0.03$           & Ly$\alpha$ &     27 $\pm$ 2 &     5000 $\pm$ 500    &   44.28 $\pm$ 0.03 \\ % s16
  $J1416+0015$ &  6.03 &   $-22.39 \pm 0.10$           & Ly$\alpha$ &     98 $\pm$ 5 &     230 $\pm$ 20       &    43.86 $\pm$ 0.01 \\ % b11
  $J1417+0117$ &  6.02 &   $-22.83 \pm 0.05$           & Ly$\alpha$ &     11 $\pm$ 1 &     420 $\pm$ 70        &    43.06 $\pm$ 0.03 \\  % a7
  $J0902+0155$ &  6.01 &   $-22.51 \pm 0.04$           & Ly$\alpha$ &     29 $\pm$ 2 &     1600 $\pm$ 1200 &    43.35 $\pm$ 0.03 \\ % a13
  $J0853+0139$ &  6.01 &   $-22.51 \pm 0.14$           & Ly$\alpha$ &     79 $\pm$ 6 &             $<$ 230        &    43.80 $\pm$ 0.01 \\ % b17
  $J1414+0130$ &  5.94 &   $-23.53 \pm 0.04$           & Ly$\alpha$ &     72 $\pm$ 3 &     2400 $\pm$ 1900 &    44.16 $\pm$ 0.01 \\ % s14
  $J0903+0211$ &  5.92 &   $-23.20 \pm 0.07$           & Ly$\alpha$ &     35 $\pm$ 5 &     1400 $\pm$ 100   &    43.70 $\pm$ 0.06 \\  % c5
$J1205-0000^*$ &  6.75 &   $-24.56 \pm 0.04$                  & ... &                   ... &                   ... &                   ... \\  % s2
$J2236+0032^*$ &   6.4 &   $-23.75 \pm 0.07$                  & ... &                   ... &                   ... &                   ... \\  % a4
$J0859+0022^*$ &  6.39 &   $-24.09 \pm 0.07$           & Ly$\alpha$ &     130 $\pm$ 5   &     540 $\pm$ 110 &    44.52 $\pm$ 0.01 \\  % c1
                            &       &                                              & NV              &     38 $\pm$ 2     &   1800 $\pm$ 200 &    43.97 $\pm$ 0.02 \\  % c1
$J1152+0055^*$ &  6.37 &   $-25.31 \pm 0.04$           & Ly$\alpha$ &     39 $\pm$ 2     & 5500 $\pm$ 1900 &    44.60 $\pm$ 0.02 \\  % s4
$J2232+0012^*$ &  6.18 &   $-22.81 \pm 0.10$           & Ly$\alpha$ &     120 $\pm$ 10 &     300 $\pm$ 30   &    44.06 $\pm$ 0.01 \\  % a3
$J2216-0016^*$ &  6.10 &   $-23.82 \pm 0.04$           & Ly$\alpha$ &     40 $\pm$ 2     &     1900 $\pm$ 300 &    44.03 $\pm$ 0.02 \\  % s1
$J2228+0128^*$ &  6.01 &   $-22.65 \pm 0.07$           & Ly$\alpha$ &     26 $\pm$ 2    &     280 $\pm$ 30    &    43.34 $\pm$ 0.02 \\  % a2
$J1207-0005^*$ &  6.01 &   $-22.77 \pm 0.06$           & Ly$\alpha$ &     8.5 $\pm$ 0.9 &     420 $\pm$ 160  &    42.92 $\pm$ 0.05 \\  % a6
$J1202-0057^*$ &  5.93 &   $-22.83 \pm 0.08$           & Ly$\alpha$ &     44 $\pm$ 6     &     1600 $\pm$ 700 &    43.66 $\pm$ 0.05 \\  % a5
  \hline
  \multicolumn{7}{c}{High-$z$ Galaxies} \\ 
  \hline
  $J1628+4312$ &  6.03 &   $-22.90 \pm 0.03$           & Ly$\alpha$ &     6.2 $\pm$ 0.3 &     230 $\pm$ 40 &    42.78 $\pm$ 0.02 \\  % a9
  $J1211-0118$ &  6.03 &   $-23.23 \pm 0.06$           & Ly$\alpha$ &     6.9 $\pm$ 0.8 &     360 $\pm$ 230 &    42.87 $\pm$ 0.04 \\ % a25
  $J1630+4315$ &  6.02 &   $-22.95 \pm 0.04$                  & ... &                   ... &                   ... &                   ... \\ % b15
  $J2233+0124$ &   6.0 &   $-22.52 \pm 0.09$                  & ... &                   ... &                   ... &                   ... \\ % b49
  $J0212-0158$ &   6.0 &   $-23.72 \pm 0.09$                  & ... &                   ... &                   ... &                   ... \\ % s12
  $J0218-0220$ &   5.9 &   $-22.94 \pm 0.04$                  & ... &                   ... &                   ... &                   ... \\ % b24
  $J0159-0359$ &  5.77 &   $-22.78 \pm 0.05$                  & ... &                   ... &                   ... &                   ... \\ % a29
  $J2237-0006$ &  5.77 &   $-22.37 \pm 0.05$                  & ... &                   ... &                   ... &                   ... \\ % b50
$J0219-0416^*$ &  5.96 &   $-22.56 \pm 0.06$                  & ... &                   ... &                   ... &                   ... \\  % b4
$J0210-0523^*$ &  5.89 &   $-23.14 \pm 0.07$                  & ... &                   ... &                   ... &                   ... \\  % b1
$J0857+0142^*$ &  5.82 &   $-22.71 \pm 0.04$           & Ly$\alpha$ &     6.2 $\pm$ 0.4 &     400 $\pm$ 50 &    42.67 $\pm$ 0.03 \\  % b6
$J0210-0559^*$ &  5.82 &   $-22.52 \pm 0.05$                  & ... &                   ... &                   ... &                   ... \\  % b2
$J0848+0045^*$ &  5.78 &   $-23.04 \pm 0.05$                  & ... &                   ... &                   ... &                   ... \\  % a1
$J0215-0555^*$ &  5.74 &   $-22.85 \pm 0.03$           & Ly$\alpha$ &     2.8 $\pm$ 0.2 &     410 $\pm$ 160 &    42.39 $\pm$ 0.04 \\  % b3
  \hline
  \multicolumn{7}{c}{[O \emissiontype{III}] Emitters} \\ 
  \hline
%  $J1157-0157$ & 0.810 &                   ...            & H$\gamma$ &     186.0 $\pm$ 28.89 &     222.8 $\pm$ 61.12 &    41.393 $\pm$ 0.038 \\ % b32
  $J1157-0157$ & 0.810 &                   ...            & H$\gamma$ &    190 $\pm$ 30 &              $<$ 190  &    41.39 $\pm$ 0.04 \\ % b32
                          &           &                                   & H$\beta$    &     340 $\pm$ 40 &             $<$ 190   &    41.65 $\pm$ 0.01 \\ % b32
                          &           &                       & [O\emissiontype{III}] $\lambda$4959 &     780 $\pm$ 100 &             $<$ 190 &    42.01 $\pm$ 0.01 \\ % b32
                         &            &                       & [O\emissiontype{III}] $\lambda$5007 &     2100 $\pm$ 300 &             $<$ 190 &    42.44 $\pm$ 0.01 \\ % b32
%  $J1443-0214$ & 0.776 &                   ...            & H$\gamma$ &     60.80 $\pm$ 11.26 &     200.6 $\pm$ 143.4 &    41.295 $\pm$ 0.070 \\ % b33
  $J1443-0214$ & 0.776 &                   ...            & H$\gamma$ &     61 $\pm$ 11 &                $<$ 190   &    41.30 $\pm$ 0.07 \\ % b33
                          &          &                                   & H$\beta$      &     130 $\pm$ 20 &             $<$ 190 &    41.64 $\pm$ 0.03 \\ % b33
                          &          &                       & [O\emissiontype{III}] $\lambda$4959 &     320 $\pm$ 30 &             $<$ 190 &    42.01 $\pm$ 0.01 \\ % b33
                          &          &                      & [O\emissiontype{III}] $\lambda$5007 &     940 $\pm$ 90 &             $<$ 190 &    42.48 $\pm$ 0.01 \\ % b33
\end{longtable}

\subsection{Quasars and Possible Quasars \label{subsec:quasars}}

We identified 24 new quasars and possible quasars at $5.9 < z \lesssim 6.8$, as displayed in Figures \ref{fig:spectra1} -- \ref{fig:spectra3} and listed in the first section of Table \ref{tab:spectroscopy}.
The highest-$z$ quasar, $J$1429$-$0104, was the only $z$-band dropout we took a spectrum of.
This quasar has two emission peaks, whose wavelengths are close to the expected positions of Ly$\alpha$ and N \emissiontype{V} $\lambda$1240 at $z = 6.8$.
The dip between the two emission lines is likely caused by a broad absorption line (BAL) system of N \emissiontype{V}.
The majority of the objects in Figures \ref{fig:spectra1} -- \ref{fig:spectra3}
exhibit characteristic spectral features of high-$z$ quasars, namely, strong and 
broad Ly$\alpha$ and in some cases N \emissiontype{V} $\lambda$1240, blue rest-UV continua, and sharp continuum breaks just shortward of Ly$\alpha$.
On the other hand, several objects have considerably narrower Ly$\alpha$ than do typical quasars.
As discussed later in this section, we classify these objects with narrow Ly$\alpha$ as (possible) quasars, given their high Ly$\alpha$ luminosity,
and possible mini BAL feature of N \emissiontype{V} found in their composite spectrum.
%We discuss the nature of these objects later in this section.

The redshifts of the above objects were determined from the Ly$\alpha$ lines, assuming that the observed line peaks correspond to the intrinsic Ly$\alpha$ wavelength 
(1216 \AA\ in the rest frame).
This assumption is not always correct, due to the strong IGM H\emissiontype{I} absorption, so the redshifts presented here (Table \ref{tab:spectroscopy}) must be interpreted with caution. 

We measured the rest-frame UV absolute magnitudes ($M_{1450}$) of these objects and the 9 quasars presented in Paper I, as follows.
All the spectra in Paper I were re-scaled to match the latest HSC photometry used in this work.
For every object, we determined the wavelength range of a continuum window, which is relatively free from strong sky emission lines; this is 9000 -- 9300 \AA\ in the most cases, while the longer wavelengths
between 9600 \AA\ and 9900 \AA\ were chosen for the objects at $z > 6.4$.
We calculated the inverse-variance-weighted mean of the flux in the continuum window, which was then extrapolated to $\lambda_{\rm rest} = 1450$ \AA,
by assuming a power-law continuum slope of $\alpha = -1.5$ ($F_{\lambda} \propto \lambda^{\alpha}$; e.g., \cite{vandenberk01}).
Since the continuum windows (corresponding to $\lambda_{\rm rest}$ = 1265 -- 1345 \AA) are close to $\lambda_{\rm rest}$ = 1450 \AA, the derived $M_{1450}$ values are not sensitive to the exact value of $\alpha$.

\begin{figure}
 \begin{center}
  \includegraphics[width=8cm]{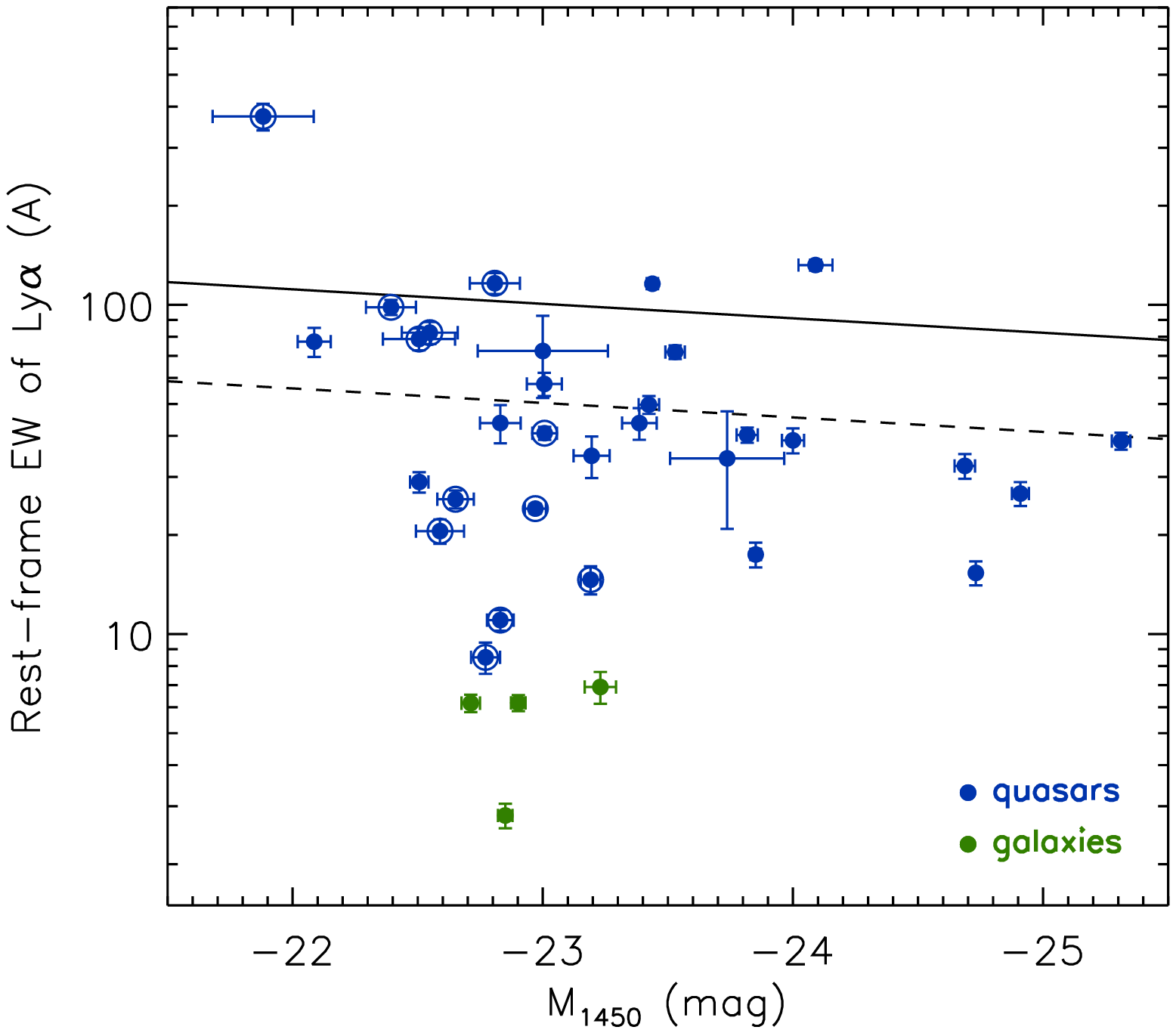} 
 \end{center}
\caption{
The UV absolute magnitudes ($M_{1450}$) and the rest-frame Ly$\alpha$ EWs of the quasars (blue dots) and the galaxies with clear Ly$\alpha$ emission (green dots).
The twelve quasars with narrow Ly$\alpha$ lines are marked with the larger circles.
The solid line represents the best-fit relation of AGNs at lower redshifts \citep{dietrich02}, while the dashed line represents the same relation modified by IGM absorption, which is assumed to absorb 
50 \% of the Ly$\alpha$ emission.
}
\label{fig:m1450-ew}
\end{figure}

We also measured the line properties (rest-frame equivalent width [EW], full width at half maximum [FWHM], and luminosity) of Ly$\alpha$ for the objects in this paper and Paper I, as follows.
For the twelve objects with narrow Ly$\alpha$ ($J0905+0300$, $J0217-0208$, $J2201+0155$, $J1423-0018$, $J1440-0107$, $J0235-0532$, $J1416+0015$, $J1417+0117$, $J0853+0139$
in this work, and $J2232+0012$, $J2228+0128$, $J1207-0005$ from Paper I), we measured the line properties with the continuum levels estimated by averaging all the pixels redward of Ly$\alpha$,
with inverse-variance weighting.
For $J0859+0022$ from Paper I, we similarly measured the above properties for the strong Ly$\alpha$ and $N\emissiontype{V}$ $\lambda$1240 lines.
For the remaining objects, we measured the properties of the  broad Ly$\alpha$ $+$ N$\emissiontype{V}$ complex, with the local continuum defined by the extrapolation of the above power-law continuum
(with the assumed slope of $\alpha = -1.5$).
Due to the difficulty in defining the accurate continuum levels, the measurements for these objects should be regarded as only approximate.
%We also measured these quantities for the nine quasars presented in Paper I, and list all the results in Table \ref{tab:spectroscopy}.
%Note that $M_{1450}$ values reported in this table are not identical to those in Paper I, due to the updated HSC photometry and the small change of the measurement method\footnote{
%In Paper I, the continuum window was fixed to the rest-frame wavelength range of $\lambda_{\rm rest} = 1270 - 1330$ \AA\ for all but the highest-$z$ object, even when the window was affected by strong sky emission lines.}.
The resultant line properties are summarized in Table \ref{tab:spectroscopy}.
Figure \ref{fig:m1450-ew} displays $M_{1450}$ and the rest-frame Ly$\alpha$ EWs of the above quasars, as well as those of the high-$z$ galaxies with clear Ly$\alpha$ emission described in \S \ref{subsec:galaxies}.
The quasars broadly follow the best-fit relation of AGNs at lower redshifts \citep{dietrich02}, if we assume IGM absorption of $\sim$50 \% of the Ly$\alpha$ emission.
%There may be a weak negative correlation between these two measures of the quasars, which is qualitatively consistent with the results at lower redshifts (e.g., \cite{dietrich02}).
A detailed analysis of the derived line properties will be presented in a future paper.

It is challenging to pin down the excitation sources of the narrow Ly$\alpha$ lines, observed in the above twelve objects. % with narrow Ly$\alpha$ is challenging with the available data.
%Meanwhile, nine objects ($J0905+0300$, $J0217-0208$, $J2201+0155$, $J1423-0018$, $J1440-0107$, $J0235-0532$, $J1416+0015$, $J1417+0117$, and $J0853+0139$) have 
%considerably smaller Ly$\alpha$ widths ($\sim$500 km s$^{-1}$) than do typical quasars.
These narrow lines are unambiguously Ly$\alpha$ at $z \sim 6$, since we observe a clear continuum break just blueward of this line in every object, which is caused by the IGM H\emissiontype{I} absorption.
%This is clearer in Figure \ref{fig:narrowLya_zoomin}, where we plot their spectra averaged in wide wavelength bins of 100 \AA.
We created a high-S/N composite spectrum by stacking the spectra of all twelve objects.
%these objects as well as three similar objects reported in Paper I ($J2232+0012$, $J2228+0128$, and $J1207-0005$).
The individual spectra were converted to the rest frame and normalized to $M_{1450} = -22.5$ mag, and then stacked together with inverse-variance weighting.
Since the twelve objects have similar continuum flux and S/N to each other, the composite spectrum has roughly equal contribution from every object in its continuum.

\begin{figure}
 \begin{center}
  \includegraphics[width=8cm]{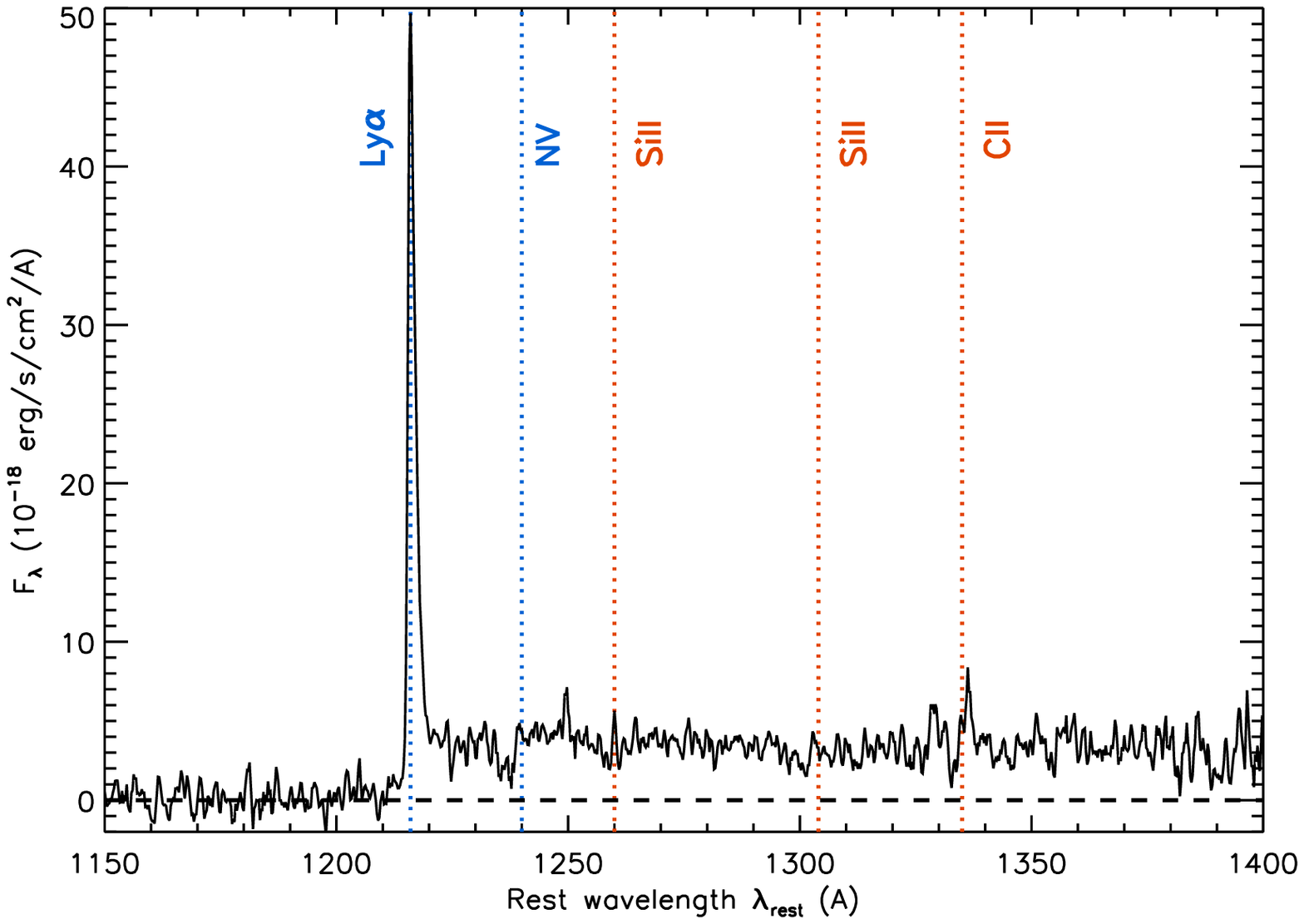} 
 \end{center}
\caption{Stacked spectrum of the twelve high-$z$ quasars with narrow Ly$\alpha$ lines, presented in the rest frame of the quasars.
The dotted lines mark the expected positions of the Ly$\alpha$, N $\emissiontype{V}$ $\lambda$1240, Si \emissiontype{II} $\lambda$1260, Si \emissiontype{II} $\lambda$1304, and C \emissiontype{II} $\lambda$1335 lines.}
\label{fig:narrowLya_stack}
\end{figure}

As displayed in Figure \ref{fig:narrowLya_stack},
we see a clear continuum break and asymmetric Ly$\alpha$ line in the composite spectrum. %, while .
The Ly$\alpha$ has a very narrow profile with a FWHM of $\sim$300 km s$^{-1}$ and the rest-frame EW of $\sim$20 \AA, but the intrinsic 
width may be as much as twice the observed value, due to IGM absorption.
On the other hand, small redshift errors will broaden the Ly$\alpha$ in the composite, so the measured FWHM is an upper limit.
For comparison, candidate type-II quasars presented by \citet{alexandroff13} have typical FWHMs of $\sim 1000$ km s$^{-1}$ in the narrow component of Ly$\alpha$.
While no N\emissiontype{V} $\lambda$1240 emission is present in the composite spectrum,
we observe a small absorption feature just blueward of the expected wavelength of N\emissiontype{V}.
This feature is not clearly seen in any of the individual spectra.
No stellar or interstellar absorption is known at this wavelength (see, e.g., the composite spectrum of Lyman break galaxies [LBGs] in \cite{shapley03}), so this feature may be a mini BAL caused by N\emissiontype{V}.
%Observations of other emission lines, such as C\emissiontype{IV} $\lambda$1549, will be useful to further probe the nature of this type of objects.
We note that a possibly similar Ly$\alpha$-only AGN, which has associated C\emissiontype{IV} $\lambda$1549 absorption systems but no metal-line emission such as N\emissiontype{V} and C\emissiontype{IV},
was found at $z = 2.5$ in SDSS \citep{hall04}.
This composite spectrum is also reminiscent of those of type-II quasars \citep{stern02, mainieri05, martinez06, alexandroff13} or radio galaxies \citep{mccarthy93}, 
which generally have very high Ly$\alpha$ EWs and little to no N\emissiontype{V} emission in their rest UV spectra.

As is evident in Figure \ref{fig:narrowLya_stack}, 
these objects have very high luminosity in the Ly $\alpha$ lines.
%In addition to the possible $N\emissiontype{V}$ mini-BAL feature, the very high luminosity of the Ly$\alpha$ lines point to the AGN nature of these objects.
%For the nine objects with narrow Ly$\alpha$, pinning down the line excitation sources is challenging with the available data.
\citet{konno16} demonstrated that, at $z \sim 2$, the majority of the Ly$\alpha$ emitters with Ly$\alpha$ luminosities $L$ (Ly$\alpha$) $\gtrsim 10^{43}$ erg s$^{-1}$ are 
%almost always 
associated with AGNs, based on their X-ray, UV, and radio properties.
Therefore, we tentatively classify all the objects with $L$ (Ly$\alpha$) $\ge 10^{43}$ erg s$^{-1}$ %or FWHM (Ly$\alpha$) $>$ 500 km s$^{-1}$ 
as possible quasars.
All the above twelve objects meet this criterion. 
We also note that the UV continuum slope of the above composite spectrum is $-1.1 \pm 0.3$, which is closer to the typical value of quasars ($\alpha = -1.5$) than to that of high-$z$ LBGs 
($\beta = -2.0$; \cite{stanway05}, \cite{bouwens14}).
Future deep observations in other wavelengths, such as X-ray, near-IR, and submm, will reveal the true nature of these intriguing sources.

\begin{figure}
 \begin{center}
  \includegraphics[width=8cm]{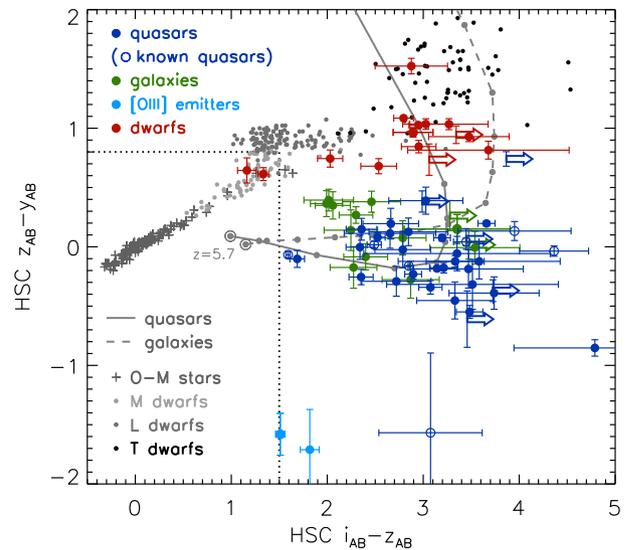} 
 \end{center}
\caption{HSC $i_{\rm AB} - z_{\rm AB}$ and $z_{\rm AB} - y_{\rm AB}$ colors of the SHELLQs quasars (blue dots), galaxies (green dots), [O \emissiontype{III}] emitters (light blue dots), brown dwarfs (red dots),
and the previously known quasars recovered in our HSC survey (blue open circles).
The grey crosses and dots represent Galactic stars and brown dwarfs, while the solid and dashed lines represent models for quasars and galaxies at $z \ge 5.7$;
the dots along the lines represent redshifts in steps of 0.1, with $z = 5.7$ marked by the large open circles.
All but the two sources at $z > 6.5$ (which have no reliable $i_{\rm AB}$ and $z_{\rm AB}$ measurements) discovered in Paper I and this work are plotted. 
}
\label{fig:colordiagram_observed}
\end{figure}

In Figure  \ref{fig:colordiagram_observed}, we plot the HSC $i_{\rm AB} - z_{\rm AB}$ and $z_{\rm AB} - y_{\rm AB}$ colors of all the spectroscopically-identified objects in Paper I and this work, as well as of 
the previously known quasars recovered in our HSC survey. %in the present HSC survey footprint.
The quasars are clearly separated from the Galactic stellar sequence on this plane.
Their colors are broadly consistent with those of the quasar model we assumed in the Bayesian algorithm, while there are outliers with 
very blue $z_{\rm AB} - y_{\rm AB}$ colors, due to exceptionally large Ly$\alpha$ EWs. % ($J0859+0022$ and $J0216-0455$).

\begin{figure}
 \begin{center}
  \includegraphics[width=8cm]{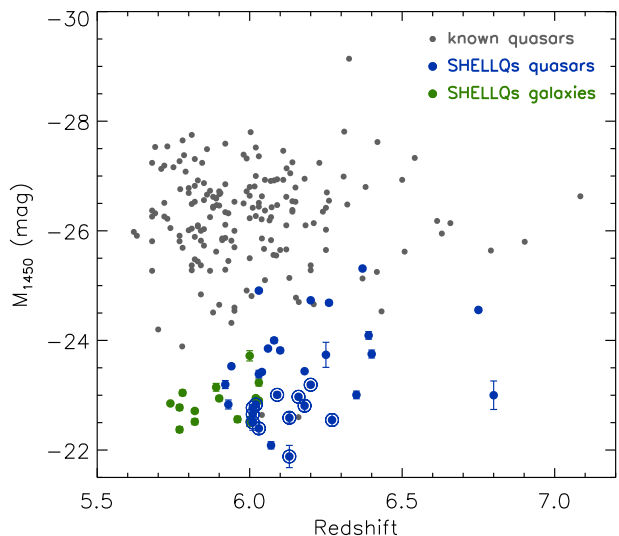} 
 \end{center}
\caption{Rest-UV absolute magnitude at 1450 \AA\ ($M_{1450}$), as a function of redshift,
	of the SHELLQs quasars (blue dots) and galaxies (green dots), as well as of all the previously known quasars in the literature (small grey dots).
	The SHELLQs quasars with narrow Ly$\alpha$ lines are marked with the larger circles.
	All the high-$z$ objects discovered in Paper I and this work are plotted.}
\label{fig:luminosity}
\end{figure}

Figure \ref{fig:luminosity} displays the absolute magnitudes $M_{1450}$ of our new quasars and galaxies (described below), along with those of high-$z$ quasars discovered previously by the other surveys.
This figure demonstrates clearly that we are opening up a new parameter space, by finding a large number of objects in the poorly-populated luminosity range of 
$M_{1450} > -25$ mag at $z > 5.7$.

\begin{figure}
 \begin{center}
  \includegraphics[width=8cm]{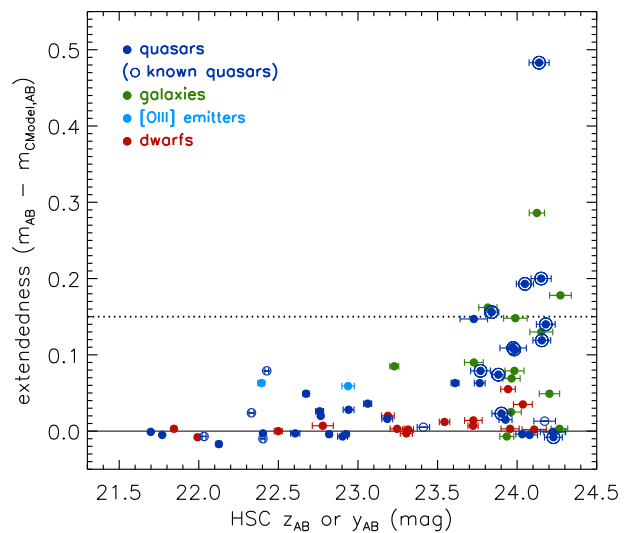} 
 \end{center}
\caption{Source extendedness defined as the difference between the PSF and CModel magnitudes, as a function of the HSC magnitudes 
	($z_{\rm AB}$ for $i$-band dropouts and $y_{\rm AB}$ for $z$-band dropouts), for the SHELLQs quasars (blue dots), galaxies (green dots), 
	[O \emissiontype{III}] emitters (light blue dots), and brown dwarfs (red dots),
	and the previously known quasars recovered in our HSC survey (blue open circles).
	The SHELLQs quasars with narrow Ly$\alpha$ lines are marked with the larger circles.
	All the sources discovered in Paper I and this work are plotted.
	The dotted line represents the extendedness cut ($m_{\rm AB} - m_{\rm CModel, AB} < 0.15$) of our quasar selection; note that some objects exceed this limit, since they 
	were selected with the older HSC photometry or with our previous selection threshold (see text).}
\label{fig:extendedness}
\end{figure}

Figure \ref{fig:extendedness} presents the source extendedness, defined as the difference between the PSF magnitudes ($m_{\rm AB}$) and CModel magnitudes ($m_{\rm CModel, AB}$).
Note that some objects exceed our extendedness cut ($m_{\rm AB} - m_{\rm CModel, AB} < 0.15$), because they had smaller extendedness %$m_{\rm AB} - m_{\rm CModel, AB}$ values 
in the older HSC data releases 
or they were selected with our previous, looser cut ($m_{\rm AB} - m_{\rm CModel, AB} < 0.30$; see Paper I).
It is notable that the objects 
%While quasars at the brighter side at $z_{\rm AB} / y_{\rm AB} \lesssim 23.5$ mag are mostly consistent with being point sources, those
at the faintest magnitudes have a long tail of the %$m_{\rm AB} - m_{\rm CModel, AB}$ 
distribution toward extended objects.
While some of these objects may have resolved host galaxies, we found that the observed distribution is consistent with that expected for simple point sources, due to photometry errors.
We test this with a special HSC dataset, which was created by stacking a part of the SSP UltraDeep data on the COSMOS field to simulate the median depth of the Wide layer.
We selected stars based on the {\it Hubble Space Telescope} ({\it HST}) Advanced Camera for Surveys (ACS) catalog \citep{leauthaud07}, and measured their HSC extendedness distribution.
We found that the fraction of stars with $z_{\rm AB} - z_{\rm CModel, AB} > 0.15$ increases toward faint magnitudes, and approaches $\sim$20 \% at $z_{\rm AB} = 24.0$ mag.
This is roughly consistent with the distribution in Figure \ref{fig:extendedness}, where four out of twenty quasars at 23.5 -- 24.5 mag have $z_{\rm AB} - z_{\rm CModel, AB} > 0.15$.
%We interpret this as evidence of the resolved host galaxies {\bf --- or is this simply due to the increasing noise at the faint end? we should compare with the similar distribution of stars}. % in these lowest-luminosity quasars ever %discovered at $z \ge 6$.
%These quasars would be excellent targets of future follow-up observations to study the host galaxies, in the optical and near-IR bands, given their lowest luminosities ever discovered 
%at $z \ge 6$; strong nuclear radiation outshining the host galaxy has been a major obstacle in such observations for luminous high-$z$ quasars.
No clear correlation is observed between the extendedness values and the PSF widths of the HSC images. %, in which the above objects are detected.

\subsection{Galaxies \label{subsec:galaxies}}

The eight objects presented in Figure \ref{fig:spectra3} have neither a broad %($> 500$ km s$^{-1}$) 
nor luminous ($> 10^{43}$ erg s$^{-1}$) Ly$\alpha$ line, %which would indicate the excitation by quasar radiation, 
and hence are most likely galaxies at $z \sim 6$.
Combined with the similar objects presented in Paper I, we have now spectroscopically identified fourteen such objects.
Their flat spectra at $\gtrsim$ 8500 \AA\ separate them clearly from Galactic brown dwarfs. % with similarly red $i - z$ colors.
As we discussed in Paper I, the steep spectral rise around $\sim$8500 \AA\ (which is responsible for the very red HSC $i_{\rm AB} - z_{\rm AB}$ colors) of these objects preclude the possibility that they 
are passive galaxies at $z \sim 1$, which would require unusually large amounts of dust extinction ($E_{\rm B-V} > 1.5$) for such type of galaxies.

The redshifts of these objects were derived from the observed positions of the Ly$\alpha$ emission, the interstellar absorptions of Si \emissiontype{II} $\lambda$1260, 
Si \emissiontype{II} $\lambda$1304, C \emissiontype{II} $\lambda$1335, and/or the continuum break caused by the IGM H \emissiontype{I} absorption.
However, this is not always easy with our spectra, due to the relatively poor S/N. %, it is hard to determine their precise redshifts.
Thus the redshifts reported here must be regarded as only approximate.

The HSC colors of these galaxies are almost indistinguishable from those of the high-$z$ quasars, because of (1) the similar intrinsic rest-UV SEDs of
the two populations and (2) the similar effect of the extrinsic IGM absorption. 
However, the weaker Ly$\alpha$ emission lines make the galaxies mildly redder than the quasars, as seen in Figure \ref{fig:colordiagram_observed}.
The galaxies may be partly resolved with the HSC angular resolution, as they have relatively large $m_{\rm AB} - m_{\rm CModel, AB}$ values (Figure \ref{fig:extendedness}).
But a similar level of extendedness is found among the faint quasars, due to photometry errors and/or resolved host galaxies as discussed in \S \ref{subsec:quasars},
so a clear distinction between these two types of objects remains very difficult with the HSC data alone.
%We will discuss this issue further in \S \ref{subsec:efficiency}

We measured the rest-frame UV absolute magnitudes $M_{1450}$ of these objects in the same way as for the quasars, assuming the continuum slope of $\beta = -2.0$ 
($F_{\lambda} \propto \lambda^{\beta}$; \cite{stanway05}).
We also measured the Ly$\alpha$ properties for the objects with Ly$\alpha$ lines detected in the spectra.
The results of these measurements are summarized in Table \ref{tab:spectroscopy}.
These high-$z$ galaxies have extremely high luminosities, in the range of $-24 \lesssim M_{1450} \lesssim -22$ mag.
They are even brighter than the galaxies identified in recent studies to constrain the bright end of the galaxy luminosity function at $z \sim 6$, which has now been 
measured at $M_{1500} \gtrsim -22.5$ mag 
(e.g., \cite{bouwens15}; \cite{bowler15}).
Therefore, the high-$z$ galaxies discovered  by our survey have the potential to provide an important clue as to the formation and evolution of most luminous galaxies in the early Universe.

\begin{figure}
 \begin{center}
  \includegraphics[width=8cm]{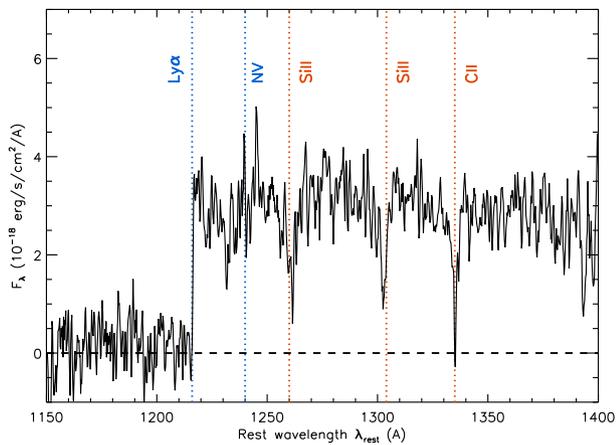} 
 \end{center}
\caption{
Stacked spectrum of the fourteen high-$z$ galaxies, presented in the rest frame of the galaxies.
The dotted lines mark the expected positions of the Ly$\alpha$, N $\emissiontype{V}$ $\lambda$1240, Si \emissiontype{II} $\lambda$1260, Si \emissiontype{II} $\lambda$1304, and C \emissiontype{II} $\lambda$1335 lines.
}
\label{fig:galaxies_composite}
\end{figure}

Figure \ref{fig:galaxies_composite} presents the composite spectrum of all the fourteen high-$z$ galaxies we discovered so far.
The individual spectra were converted to the rest frame and normalized to $M_{1450} = -22.5$ mag, and then stacked together with inverse-variance weighting.
Since redshifts of the galaxies, those without Ly$\alpha$ lines in particular, cannot be determined accurately, any spectral features of the individual spectra are smeared in this composite.
Nonetheless, we detected strong absorption lines of Si \emissiontype{II} $\lambda$1260, Si \emissiontype{II} $\lambda$1304, and C \emissiontype{II} $\lambda$1335, whose rest-frame EWs are
2.0 $\pm$ 0.5 \AA, 1.3 $\pm$ 0.4 \AA, 2.3 $\pm$ 0.4 \AA, respectively.
These are broadly consistent with the EWs measured in the composite spectrum of $z \sim 3$ LBGs presented by \citet{shapley03}.
The UV spectral slope of our composite spectrum is $\beta = -1.8 \pm 0.1$, which is close to the commonly-assumed value of $\beta = -2.0$ \citep{stanway05}.

We note that there is a separate project to search for high-$z$ galaxies from the HSC-SSP dataset \citep{ono17}.
They apply no extendedness cut in the sample selection, which is complementary to our quasar selection.
In addition, \citet{shibuya17} are carrying out spectroscopic observations of the HSC sources with excess brightness in narrow-band filters, in the SSP Deep and UltraDeep fields.
They have already identified 21 Ly$\alpha$ emitters (LAEs) at $z$ = 6 -- 7, with high Ly$\alpha$ luminosities and equivalent widths. 
These LAEs may partly overlap with the population of possible quasars with narrow Ly$\alpha$ discussed above, but a detailed cross comparison is beyond the scope of this paper.
Combining the present results with the above works will provide a more complete census of the high-$z$ Universe, covering a wide range of galaxy properties.

\subsection{[O \emissiontype{III}] Emitters}

Unexpectedly, we identified two strong [O \emissiontype{III}] emitters among the quasar candidates, as presented in Figures \ref{fig:spectra5}.
We measured the line properties of H$\gamma$, H$\beta$, [O \emissiontype{III}] $\lambda$4959 and $\lambda$5007 of these objects, and listed the results in Table \ref{tab:spectroscopy}.
Since they have very weak continua, we estimated the continuum levels by summing up all the available pixels except for the above emission lines. % to estimate the continuum levels.
Their very high [O \emissiontype{III}] $\lambda$5007/H$\beta$ ratios (log ([O \emissiontype{III}] $\lambda$5007/H$\beta$) $\sim 0.8$) are achievable only 
in galaxies with sub-solar metallicity and high ionization state of the interstellar medium (e.g., \cite{kewley16}).
Since the theoretical prediction of \citet{kewley16} doesn't exceed log ([O \emissiontype{III}] $\lambda$5007/H$\beta$) = 0.7 in all the assumed cases,
there may be AGN contribution in our [O \emissiontype{III}] emitters.
Alternatively, they may be extreme emission-line galaxies with compact morphology and low metallicity, which are know to exhibit unusually high [O \emissiontype{III}] EWs 
and [O \emissiontype{III}]/H$\beta$ ratios \citep{amorin14, amorin15}.
The H$\gamma$/H$\beta$ ratios of these objects are close to the intrinsic value of 0.47 \citep{osterbrock06}, which indicates that there is little dust extinction.

\subsection{Brown Dwarfs}

We found 14 new brown dwarfs, as presented in Figures \ref{fig:spectra6} and \ref{fig:spectra7}.
We derived their spectral classes by fitting the spectral standard templates of M4- to T8-type dwarfs, taken from the SpeX Prism Spectral Library \citep{burgasser14, skrzypek15}, to
the observed spectra at $\lambda_{\rm obs} = 7500 - 9800$ \AA.
The results are summarized in Table \ref{tab:bdtypes} and plotted in Figures \ref{fig:spectra6} and \ref{fig:spectra7}.
Due to the relatively low S/N and limited spectral coverage, we regard the spectral classes presented here as only approximate.
Table \ref{tab:bdtypes} also reports the spectral class of $J0850+0012$, the one brown dwarf presented in Paper I.
These brown dwarfs are of the late-M to T types, which are exactly what we assumed as major contaminants in the quasar selection algorithm.

We note that there are other HSC-SSP projects to look for faint brown dwarfs (Chiang et al., in prep.; Sorahana et al., in prep).
%, which exploit the unprecedented depth of the survey over the wide area.
Combined with the results from these projects, the present brown dwarfs will provide important clues as to the nature of the Galaxy, such as star-formation history, 
initial mass function, and spatial structure.

\begin{table}
  \tbl{Spectral classes of the brown dwarfs.}{%
  \begin{tabular}{ll}
      \hline
      Name & Class \\ 
      \hline
  $J0210-0451$ & L9 \\ 
  $J0211-0414$ & T1 \\ 
  $J0214-0214$ & T2 \\ 
  $J0214-0645$ & L9 \\ 
  $J0217-0708$ & M7 \\ 
  $J0226-0403$ & M7 \\ 
  $J0230-0623$ & L9 \\ 
  $J0234-0604$ & T2 \\ 
  $J0854-0004$ & T5 \\ 
  $J1204-0046$ & T0 \\ 
  $J2206+0231$ & L9 \\ 
  $J2209+0139$ & T1 \\ 
  $J2211-0027$ & L7\\ 
  $J2237+0239$ &  L9 \\ 
$J0850+0012^*$ & T1 \\ 
      \hline
    \end{tabular}}\label{tab:bdtypes}
\begin{tabnote}
$^*$This object is taken from Paper I.
\end{tabnote}
\end{table}

\subsection{Survey Efficiency \label{subsec:efficiency}}

The efficiency of our high-$z$ quasar survey remains quite high. 
We have identified the nature of 64 HSC sources in Paper I and this work, which include 33 high-$z$ quasars, 14 high-$z$ galaxies, 2 [O\emissiontype{III}] emitters,
and 15 brown dwarfs.
In addition to the above objects, we took follow-up images or spectra of 13 quasar candidates, but they were not detected for unknown reasons.
Since they should have been detected with our exposure times based on the HSC magnitudes, they are most likely transient or moving sources. %, such as supernovae and asteroids.
We are still investigating what these sources could be, and will present the results in a forthcoming paper.

\begin{figure}
 \begin{center}
  \includegraphics[width=8cm]{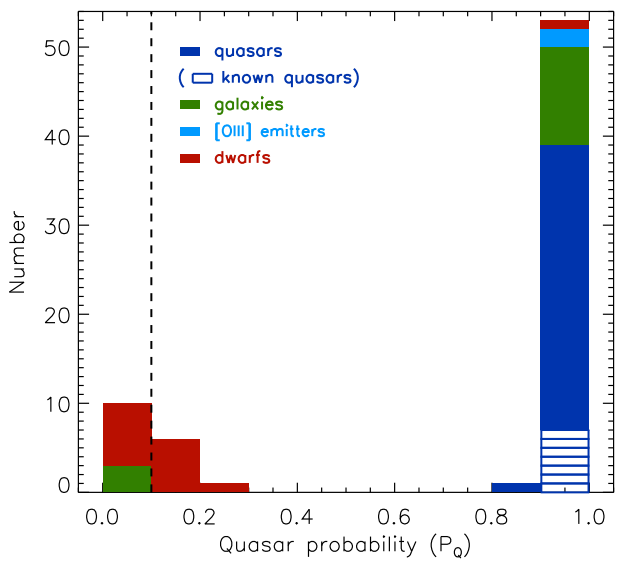} 
 \end{center}
\caption{Histogram of the Bayesian quasar probability ($P_{\rm Q}^{\rm B}$) of the SHELLQs quasars (blue), galaxies (green), [O \emissiontype{III}] emitters (light blue), brown dwarfs (red),
and the previously known quasars recovered in our HSC survey (white with blue outline).
All the sources discovered in Paper I and this work are counted.
The dashed line represents our quasar selection threshold ($P_{\rm Q}^{\rm B} > 0.1$); note that several objects have lower $P_{\rm Q}^{\rm B}$ values than this threshold, because
they were selected with older HSC data releases, which indicated higher $P_{\rm Q}^{\rm B}$ values.}
\label{fig:Pq}
\end{figure}

Figure \ref{fig:Pq} displays a histogram of the Bayesian quasar probability ($P_{\rm Q}^{\rm B}$), for all the spectroscopically identified objects in Paper I and this work.
%Moreover, we found that most of the contaminating brown dwarfs have low values of the Bayesian quasar probability $P_{\rm Q}^{\rm B}$. % calculated with the HSC and matched near-IR photometry.
As we mentioned previously, the $P_{\rm Q}^{\rm B}$ values have a clear bimodal distribution.
The peak at around $P_{\rm Q}^{\rm B} = 0.0$ is populated mostly by brown dwarfs, which means that we knew, before spectroscopy, that these HSC sources were not very promising quasar candidates. 
Many of these dwarfs have lower $P_{\rm Q}^{\rm B}$ values than our quasar selection threshold ($P_{\rm Q}^{\rm B} = 0.1$), due to the improved HSC photometry with the new data
reduction pipeline; we took their spectra because they had $P_{\rm Q}^{\rm B} > 0.1$ in the older data releases.

The other peak of the distribution at around $P_{\rm Q}^{\rm B} = 1.0$ is populated mostly by high-$z$ quasars. 
All the discovered quasars have $P_{\rm Q}^{\rm B} > 0.8$, which implies that their spectral diversity is reasonably covered by the quasar model in our Bayesian probabilistic algorithm.
Figure \ref{fig:Pq} suggests that we could further improve the success rate of quasar discovery, by raising the selection threshold to, e.g., $P_{\rm Q}^{\rm B} = 0.5$.
However, we will keep the present threshold for the time being, as we continue the survey, so that we do not miss any quasars with unusual (and thus potentially interesting) spectral properties.

%So far, there is only a limited advantage of including UKIDSS and VIKING photometry to our quasar selection.
%Six of the above 64 objects are detected in $J$, $H$, and/or $K$ band, as summarized in Table \ref{tab:nir_photometry}.
%We found that in all but $J0854-0004$ case, the Bayesian quasar probability doesn't change much with and without the NIR magnitudes;
%with the HSC data alone, $P_{\rm Q}^{\rm B}$ is close to unity for the two quasars, and close to zero for the three brown dwarfs.
%For $J0854-0004$, $P_{\rm Q}^{\rm B}$ changes from 0.61 to 0.00 by including the VIKING magnitudes, because the bright $JH$ magnitudes of this object (which has $z_{\rm AB}$ = 23.544 $\pm$ 0.033 mag
%from Table \ref{tab:photometry}) indicate a very red SED, precluding the possibility of being a high-$z$ quasar.
%In addition to the above objects, 27 objects fall on the VIKING survey footprint but are not detected in any of the $JHK$ bands.
%Although we used flux upper limits in these non-detection cases in the Bayesian algorithm, the VIKING depth was too shallow to impact the resultant quasar probability
%for all these HSC sources.

Figure \ref{fig:extendedness} demonstrates that our present extendedness cut ($m_{\rm AB} - m_{\rm CModel, AB} < 0.15$) does not recover all the possible quasars in the HSC survey.
Four out of the twenty quasars with 23.5 -- 24.5 mag have larger extendedness, which is consistent with the %$m_{\rm AB} - m_{\rm CModel, AB}$ 
distribution of the {\it HST}/ACS stars, as described 
in \S \ref{subsec:quasars}. % ($\sim$20 \% of the {\it HST}/ACS stars have $z_{\rm PSF} - z_{\rm CModel} > 0.15$ at $z_{\rm PSF} = 24.0$ mag).
The exact value of this extendedness cut should be defined as a compromise between completeness and purity of quasar selection, given the available amount of 
telescope time for spectroscopic identification, and we think the present cut is a reasonable choice for our survey.
Of course, the above $\sim$20 \% loss of point sources due to the extendedness cut should be corrected for when we measure the luminosity function and other statistical properties.

Contamination by high-$z$ galaxies significantly reduces the purity of quasars among the photometric candidates at the faintest magnitudes (e.g., Figure \ref{fig:extendedness}).
As we discussed in \S \ref{subsec:galaxies}, it is difficult in practice to distinguish high-$z$ quasars and galaxies with the HSC photometry data only.
However, those bright galaxies are a very useful probe of the Universe in the reionization era. 
They are excellent targets to study stellar populations at high redshift, with deep optical and near-IR follow-up observations.
We also plan to measure their gas and dust properties using, e.g., the Atacama Large Millimeter and sub-millimeter Array (ALMA).
We will continue to produce a sample of such galaxies from our survey at $z_{\rm AB} \lesssim 24.5$ mag; at fainter magnitudes, galaxies outnumber quasars and
we would need prohibitively large amounts of telescope time to find quasars among the large number of galaxy targets.

\subsection{Gravitational Lensing}

Given the high probability of gravitational lensing magnification for high-$z$ objects (e.g., \cite{wyithe11}), it is worthwhile to check the possibility of lensing magnifications for 
the sample of objects presented in this paper. 
We did so by cross-correlating our high-$z$ quasars and galaxies with potential foreground deflectors, such as massive galaxies and clusters of galaxies. 
In the present work, we used two catalogs of HSC foreground objects available at $z < 1.1$.
First, we used an HSC cluster catalog with richness $N > 15$ and in the redshift interval $0.1 < z < 1.1$ \citep{oguri17}, and found that there are no matches within 60\arcsec. 
Second, we used a photometric luminous red galaxy sample with stellar mass $M_* > 10^{10.3} M_\odot$ and at redshift $0.05 < z < 1.05$ (Oguri et al., in prep.), 
and found 7 matches within $10''$. 
We estimated magnifications by these nearby red galaxies by converting the stellar masses to stellar velocity dispersions, using the scaling relation derived in SDSS \citep{kauffmann03}, 
and adopting a singular isothermal sphere for the mass distribution of the individual galaxies.
We found that the magnification factors by these galaxies are small, $\mu \lesssim 1.2$ at most. 
A potentially interesting object is the galaxy $J2233+0124$; there is  a red galaxy at $z\sim 0.66$ with an angular separation of 1\arcsec.9, %from $J2233+0124$, 
and there is also a cluster of galaxies with richness $N \sim 15$ at the similar redshift, $z\sim 0.65$, at a separation of $\sim$ 90\arcsec. % from $J2233+0124$.
However, the magnification by these foreground sources %this nearby galaxy or cluster of galaxies 
is estimated to be small, assuming the standard scaling relations between galaxy properties and underlying mass distribution.

Although the above analysis does not test all the potential deflectors, we conclude for now that there is no evidence of lensing magnification at work 
in the present sample of high-$z$ quasars and galaxies.
A more effective test will be to observe these objects with higher angular resolution, e.g., with the {\it HST} or the {\it James Webb Space Telescope}, to look for
lensed morphology or companion.
%\citet{richards04} and \citet{richards06}
We note that Richards et al. (2004,2006) carried out a {\it HST}/ACS snapshot survey of SDSS quasars, but found no case of strong lensing in 161 quasars at $4.0 < z < 6.4$.

\section{Summary\label{sec:summary}}

This paper is the second in a series presenting the results of the SHELLQs project, a survey of low-luminosity quasars at high redshift ($z > 5.7$) close to the reionization era.
Quasar candidates are selected with a Bayesian probabilisitic algorithm, using the multi-band imaging data of the Subaru HSC-SSP survey.
We took optical spectra of 48 candidates with GTC/OSIRIS and Subaru/FOCAS, and newly discovered 24 quasars and 8 luminous galaxies at $5.7 < z \le 6.8$.
Combined with the sample presented in Paper I, we have now identified 64 HSC sources over about 430 deg$^2$, which include 33 high-$z$ quasars, 
14 high-$z$ luminous galaxies, 2 [O\emissiontype{III}] emitters at $z \sim 0.8$, and 15 Galactic brown dwarfs.
We present a spectral analysis of all these objects in this paper.

The new quasars have considerably lower luminosity ($M_{1450} \sim -25$ to $-22$ mag) than most of the previously known high-$z$ quasars.
Several of these quasars have luminous ($> 10^{43}$ erg s$^{-1}$) and narrow ($< 500$ km s$^{-1}$) Ly$\alpha$ lines, and also a possible mini BAL system of N\emissiontype{V} $\lambda$1240 
in the composite spectrum, which clearly separate them from typical quasars.
On the other hand, the high-$z$ galaxies have extremely high luminosity ($M_{1450} \sim -24$ to $-22$ mag) compared to other galaxies found at similar redshift.
With the discovery of these new classes of objects, we are opening up new parameter spaces in the high-$z$ Universe.
The two [O\emissiontype{III}] emitters are likely to be low-metallicity star-forming galaxies at $z \sim 0.8$, but there may be AGN contribution to the very strong [O\emissiontype{III}] lines.
The brown dwarfs have the spectral classes from late-M to T, which are exactly what we assumed as major contaminants in the quasar selection algorithm.
Our survey remains quite efficient, with most of the objects with significant Bayesian quasar probability ($P_{\rm Q}^{\rm B} > 0.5$) being identified as high-$z$ quasars or galaxies spectroscopically.

The SHELLQs project will continue, as the HSC-SSP survey continues toward its goals of observing 1400 deg$^2$ in the Wide layer,
as well as 27 and 3.5 deg$^2$ in the Deep and UltraDeep layer, respectively.
%As we described in Paper I, from the entire Wide field, 
We expect to discover $\sim$500 quasars with $z_{\rm AB} < 24.5$ mag at $z \sim 6$, and $\sim$100 quasars with $y_{\rm AB} < 24.0$ mag at $z \sim 7$,
in the entire Wide field (see Paper I).
No other survey has an ability to find such a large sample of high-$z$ low-luminosity quasars, before the advent of Large Synoptic Survey Telescope.
%while deep survey components of Dark Energy Survey and Pan-STARRS1 survey, or other drilling surveys over $>10$ deg$^2$
We are discovering more new quasars while this paper is written, which will be reported in forthcoming papers.
We will also derive our first quasar luminosity function at $z \sim 6$, reaching down to $M_{\rm AB} \sim -22$ mag, very soon.
Follow-up observations of the discovered objects are being considered at various wavelengths from sub-millimeter/radio to X-ray.
Several of them have already been observed with ALMA and near-IR spectrographs on the Gemini telescope and Very Large Telescope, whose results will be presented elsewhere.

%%%%%%%%%%%%%%%%%%%%%%%%%%%%%%%%%%%%%%%

%\begin{table}
%  \tbl{This is the first tabular.}{%
%  \begin{tabular}{llll}
%      \hline
%      Name & Value1 & Value2 \\ 
%      \hline
%      aaa & bbb & ccc & ddd \\
%      eee & fff & ggg & hhh \\
%      ....\\
%      \hline
%    \end{tabular}}\label{tab:first}
%\begin{tabnote}
%This is table note.
%\end{tabnote}
%\end{table}

%%%%%%%%%%%%%%%%%%%%%%%%%%%%%%%%%%%%%%%

\begin{ack}

This work is based on data collected at the Subaru Telescope, which is operated by the National Astronomical Observatory of Japan (NAOJ).
We appreciate the staff members of the telescope for their support during our FOCAS observations.
The data analysis was in part carried out on the open use data analysis computer system at the Astronomy Data Center of NAOJ.

This work is also based on observations made with the Gran Telescopio Canarias (GTC), installed at the Spanish Observatorio del Roque de los Muchachos 
of the Instituto de Astrof\'{i}sica de Canarias, on the island of La Palma.
We thank Stefan Geier and other support astronomers for their help during preparation and execution of our observing program.

YM was supported by JSPS KAKENHI Grant No. JP17H04830.
NK acknowledges support from the JSPS through Grant-in-Aid for Scientific Research 15H03645.
KI acknowledges support by the Spanish MINECO under grant AYA2016-76012-C3-1-P and
MDM-2014-0369 of ICCUB (Unidad de Excelencia 'Mar\'ia de Maeztu').
TN acknowledges support from the JSPS (KAKENHI grant no. 16H01101 and 16H03958).
KK was supported by JSPS Grant-in-Aid for Scientific Research (A) Number 25247019.

The Hyper Suprime-Cam (HSC) collaboration includes the astronomical
communities of Japan and Taiwan, and Princeton University.  The HSC
instrumentation and software were developed by NAOJ, the Kavli Institute for the
Physics and Mathematics of the Universe (Kavli IPMU), the University
of Tokyo, the High Energy Accelerator Research Organization (KEK), the
Academia Sinica Institute for Astronomy and Astrophysics in Taiwan
(ASIAA), and Princeton University.  Funding was contributed by the FIRST 
program from Japanese Cabinet Office, the Ministry of Education, Culture, 
Sports, Science and Technology (MEXT), the Japan Society for the 
Promotion of Science (JSPS),  Japan Science and Technology Agency 
(JST),  the Toray Science  Foundation, NAOJ, Kavli IPMU, KEK, ASIAA,  
and Princeton University.

This paper makes use of software developed for the Large Synoptic Survey Telescope (LSST). We thank the LSST Project for 
making their code available as free software at http://dm.lsst.org.

The Pan-STARRS1 Surveys (PS1) have been made possible through contributions of the Institute for Astronomy, the University of Hawaii, the Pan-STARRS Project Office, the Max-Planck Society and its participating institutes, the Max Planck Institute for Astronomy, Heidelberg and the Max Planck Institute for Extraterrestrial Physics, Garching, The Johns Hopkins University, Durham University, the University of Edinburgh, Queen's University Belfast, the Harvard-Smithsonian Center for Astrophysics, the Las Cumbres Observatory Global Telescope Network Incorporated, the National Central University of Taiwan, the Space Telescope Science Institute, the National Aeronautics and Space Administration under Grant No. NNX08AR22G issued through the Planetary Science Division of the NASA Science Mission Directorate, the National Science Foundation under Grant No. AST-1238877, the University of Maryland, Eotvos Lorand University (ELTE) and the Los Alamos National Laboratory.

IRAF is distributed by the National 
Optical Astronomy Observatory, which is operated by the Association of Universities for Research in Astronomy (AURA) under a cooperative agreement 
with the National Science Foundation.

\end{ack}

%\appendix 
%\section*{Case of single paragraph}
%\section{Case of two or paragraphs}
%\section{Case of two or paragraphs}

%%%
% See the manual for the detail.
%%%

\end{document}